\documentclass[epj,twocolumn,superscriptaddress,nofootinbib]{revtex4}

\usepackage{amsmath,amsfonts,amssymb,bm}
\usepackage{dsfont}
\usepackage{enumitem}
\usepackage{graphicx}
\usepackage{color}
\usepackage{soul}

\usepackage[mathscr,scaled=1.15]{urwchancal}
\DeclareFontFamily{OT1}{pzc}{}
\DeclareFontShape{OT1}{pzc}{m}{it}%
{<-> s * [1.15] pzcmi7t}{}
\DeclareMathAlphabet{\mathpzc}{OT1}{pzc}{m}{it}

\definecolor{purple}{rgb}{0.5,0,0.5}
\definecolor{blue}{rgb}{0.0,0,0.9}
\definecolor{prdblue}{rgb}{0.133,0.118,0.498}
\usepackage[colorlinks=true, pdfstartview=FitV, linkcolor=prdblue, citecolor= prdblue, urlcolor=prdblue]{hyperref}

\begin{document}

\title{%%$\,$\\[-7ex]\hspace*{\fill}{\small{\emph{Preprint no}. NJU-INP 001/19}}\\[1ex]
Pion and Kaon Structure at the Electron-Ion Collider}
%\subtitle{Do you have a subtitle?\\ If so, write it here}

%\titlerunning{Short form of title}        % if too long for running head

\author{Arlene C. Aguilar}
\affiliation{
University of Campinas - UNICAMP, Institute of Physics ``Gled Wataghin'',
13083-859 Campinas, S\~ao Paulo, Brazil}

\author{Zafir Ahmed}
%\email{zahmed@jlab.org}
\affiliation{University of Regina, Regina, Saskatchewan, S4S 0A2, Canada}

\author{Christine Aidala}
%\email[]{caidala@umich.edu}
\affiliation{University of Michigan, Ann Arbor, Michigan 48109-1040, USA}
%% Grant?

\author{Salina Ali}
%\email{95ali@cua.edu}
\affiliation{Catholic University of America, Washington DC 20064, USA}

\author{Vincent Andrieux}
\affiliation{University of Illinois at Urbana-Champaign, Urbana, Illinois 61801, USA}
\affiliation{CERN, 1211 Geneva 23, Switzerland}

\author{John Arrington}
\affiliation{Argonne National Laboratory, Lemont, IL 60439, USA}

\author{Adnan Bashir}
\affiliation{Instituto de F\'{\i}sica y Matem\'aticas, Universidad
Michoacana de San Nicol\'as de Hidalgo\\
Edificio C-3, Ciudad Universitaria, C.P. 58040,
Morelia, Michoac\'an, M{\'e}xico}

\author{Vladimir Berdnikov}
%\email{berdnik@jlab.org}
\affiliation{Catholic University of America, Washington DC 20064, USA}

\author{Daniele Binosi}
\affiliation{European Centre for Theoretical Studies in Nuclear Physics
and Related Areas (ECT$^\ast$) and Fondazione Bruno Kessler\\ Villa Tambosi, Strada delle Tabarelle 286, I-38123 Villazzano (TN) Italy}

\author{Lei Chang}
%\email[]{leichang@nankai.edu.cn}
\affiliation{School of Physics, Nankai University, Tianjin 300071, China}

\author{Chen Chen}
%\email[]{Chen.Chen@theo.physik.uni-giessen.de}
\affiliation{Institut f\"ur Theoretische Physik, Justus-Liebig-Universit\"at Gie{\ss}en, 35392 Gie{\ss}en, Germany}

\author{Muyang Chen}
%\email[]{leichang@nankai.edu.cn}
\affiliation{School of Physics, Nankai University, Tianjin 300071, China}

\author{Jo\~ao Pacheco B. C. de Melo}
%\email{jpachecodm@gmail.com}
%\affiliation{Universidade Cruzeiro do Sul, Sao Paulo, Brazil}
\affiliation{Laborat\'{o}rio de F\'{\i}sica Te\'{o}rica e Computacional -- LFTC,
  Universidade Cruzeiro do Sul / Universidade Cidade de S\~{a}o Paulo,
  01506-000 S\~{a}o Paulo, SP, Brazil}

\author{Markus Diefenthaler}
%\email[]{mdiefent@jlab.gov}
\affiliation{Thomas Jefferson National Accelerator Facility, Newport News, VA 23606, USA}

\author{Minghui Ding}
\affiliation{European Centre for Theoretical Studies in Nuclear Physics
and Related Areas (ECT$^\ast$) and Fondazione Bruno Kessler\\ Villa Tambosi, Strada delle Tabarelle 286, I-38123 Villazzano (TN) Italy}
\affiliation{School of Physics, Nankai University, Tianjin 300071, China}

\author{Rolf Ent}
%\email[]{ent@jlab.org}
\affiliation{Thomas Jefferson National Accelerator Facility, Newport News, VA 23606, USA}

\author{Tobias Frederico}
%\email{tobias@ita.br}
\affiliation{Instituto Tecnol\'ogico de Aeron\'autica, 12.228-900 S\~ao Jos\'e dos Campos, Brazil}

\author{Fei Gao}
\affiliation{Institut f\"ur Theoretische Physik, Universit\"at Heidelberg,
Philosophenweg 16, 69120 Heidelberg, Germany}

\author{Ralf W. Gothe}
\affiliation{University of South Carolina, Columbia, SC 29208, USA}

\author{Mohammad Hattawy}
%\email[]{mhattawy@odu.edu}
\affiliation{Old Dominion University, Norfolk, Virginia 23529, USA}

\author{Timothy J. Hobbs}
%\email{tjhobbs@smu.edu}
\affiliation{Southern Methodist University, Dallas, TX 75275-0175, USA}

\author{Tanja Horn}
%\email[]{hornt@cua.edu}
\affiliation{Catholic University of America, Washington DC 20064, USA}

\author{Garth M. Huber}
%\email[]{huberg@uregina.ca}
\affiliation{University of Regina, Regina, Saskatchewan, S4S 0A2, Canada}

\author{Shaoyang Jia}
%\email[]{sjia@iastate.edu}
\affiliation{Iowa State University, Ames, IA 50011, USA}

\author{Cynthia Keppel}
%\email[]{keppel@jlab.org}
\affiliation{Thomas Jefferson National Accelerator Facility, Newport News, VA 23606, USA}

\author{Gast\~ao Krein}
%\email[]{gkrein@ift.unesp.br}
\affiliation{Instituto de F\'isica Te\'orica, Universidade Estadual Paulista, Rua Dr.~Bento Teobaldo Ferraz, 271, 01140-070 S\~ao Paulo, SP, Brazil}

\author{Huey-Wen Lin}
%\email[]{hwlin@pa.msu.edu}
\affiliation{Michigan State University, East Lansing, Michigan, 48824, USA}

\author{C\'edric Mezrag}
\affiliation{Istituto Nazionale di Fisica Nucleare, Sezione di Roma,
P. le A. Moro 2, I-00185 Roma, Italy}

\author{Victor Mokeev}
\affiliation{Thomas Jefferson National Accelerator Facility, Newport News, VA 23606, USA}

\author{Rachel Montgomery}
%\email[]{Rachel.Montgomery@glasgow.ac.uk}
\affiliation{University of Glasgow, G128QQ Scotland, UK}

\author{Herv\'e Moutarde}
\affiliation{IRFU, CEA, Universit\'e Paris-Saclay, F-91191 Gif-sur-Yvette, France}

\author{Pavel Nadolsky}
%\email{nadolsky@physics.smu.edu}
\affiliation{Southern Methodist University, Dallas, TX 75275-0175, USA}

\author{Joannis Papavassiliou}
\affiliation{Department of Theoretical Physics and IFIC, University of Valencia and CSIC, E-46100, Valencia, Spain}

\author{Kijun Park}
%\email[]{parkkj@jlab.org}
\affiliation{Thomas Jefferson National Accelerator Facility, Newport News, VA 23606, USA}

\author{Ian L. Pegg}
%\email[]{ianp@vsl.cua.edu}
\affiliation{Catholic University of America, Washington DC 20064, USA}

\author{Jen-Chieh Peng}
%\email[]{jcpeng@illinois.edu}
\affiliation{University of Illinois at Urbana-Champaign, Urbana, Illinois 61801, USA}

\author{Stephane Platchkov}
%\email[]{stephane.platchkov@cea.fr}
\affiliation{IRFU, CEA, Universit\'e Paris-Saclay, F-91191 Gif-sur-Yvette, France}

\author{Si-Xue Qin}
\affiliation{Department of Physics, Chongqing University, Chongqing 401331, P.\,R. China}

\author{Kh\'epani Raya}
\affiliation{School of Physics, Nankai University, Tianjin 300071, China}

\author{Paul Reimer}
%\email[]{reimer@anl.gov}
\affiliation{Argonne National Laboratory, Lemont, IL 60439, USA}

\author{David G. Richards}
%\email[]{keppel@jlab.org}
\affiliation{Thomas Jefferson National Accelerator Facility, Newport News, VA 23606, USA}

\author{Craig D. Roberts}
%\email[]{cdroberts@nju.edu.cn}
\affiliation{Department of Physics, Nanjing University, Nanjing, Jiangsu 210093, China}
\affiliation{Institute for Nonperturbative Physics, Nanjing University, Nanjing, Jiangsu 210093, China}

\author{Jose Rodr\'{\i}guez-Quintero}
\affiliation{Department of Integrated Sciences and Centre for Advanced Studies in Physics,
Mathematics and Computation, University of Huelva, E-21071 Huelva, Spain}

\author{Nobuo Sato}
%\email[]{nsato@jlab.org}
\affiliation{Thomas Jefferson National Accelerator Facility, Newport News, VA 23606, USA}

\author{Sebastian M. Schmidt}
%\email[]{s.schmidt@fz-juelich.de}
\affiliation{
Institute for Advanced Simulation, Forschungszentrum J\"ulich and JARA, D-52425 J\"ulich, Germany}

\author{Jorge Segovia}
\affiliation{Departamento de Sistemas F\'isicos, Qu\'imicos y Naturales, Universidad Pablo de Olavide, E-41013 Sevilla, Spain}

\author{Arun Tadepalli}
%\email[]{arunts@jlab.org}
\affiliation{Thomas Jefferson National Accelerator Facility, Newport News, VA 23606, USA}

\author{Richard Trotta}
%\email[]{trotta@cua.edu}
\affiliation{Catholic University of America, Washington DC 20064, USA}

\author{Zhihong Ye}
%\email[]{yez@anl.gov}
\affiliation{Argonne National Laboratory, Lemont, IL 60439, USA}

\author{Rikutaro Yoshida}
%\email[]{ryoshida@jlab.org}
\affiliation{Thomas Jefferson National Accelerator Facility, Newport News, VA 23606, USA}

\author{Shu-Sheng Xu}
\affiliation{College of Science, Nanjing University of Posts and Telecommunications, Nanjing 210023, China}

%% 51 signatures

\date{16 September 2019}
%\date{17 July 2019}
%\date{08 June 2019}

\begin{abstract}
\hspace*{-\parindent}\mbox{\sf Abstract}.
Understanding the origin and dynamics of hadron structure and in turn that of atomic nuclei is a central goal of nuclear physics. This challenge entails the questions of how does the roughly $1$ GeV mass-scale that characterizes atomic nuclei appear; why does it have the observed value; and, enigmatically, why are the composite Nambu-Goldstone (NG) bosons in quantum chromodynamics (QCD) abnormally light in comparison? In this perspective, we provide an analysis of the mass budget of the pion and proton in QCD; discuss the special role of the kaon, which lies near the boundary between dominance of strong and Higgs mass-generation mechanisms; and explain the need for a coherent effort in QCD phenomenology and continuum calculations, in exa-scale computing as provided by lattice QCD, and in experiments to make progress in understanding the origins of hadron masses and the distribution of that mass within them. We compare the unique capabilities foreseen at the electron-ion collider (EIC) with those at the hadron-electron ring accelerator (HERA), the only previous electron-proton collider; and describe five key experimental measurements, enabled by the EIC and aimed at delivering fundamental insights that will generate concrete answers to the questions of how mass and structure arise in the pion and kaon, the Standard Model's NG modes, whose surprisingly low mass is critical to the evolution of our Universe.\\[1ex]
\emph{Corresponding authors}:\\
Rolf Ent (\href{mailto:ent@jlab.org}{ent@jlab.org});
Tanja Horn (\href{mailto:hornt@cua.edu}{hornt@cua.edu});
Craig Roberts (\href{mailto:cdroberts@nju.edu.cn}{cdroberts@nju.edu.cn}); and
Rikutaro Yoshida (\href{mailto:ryoshida@jlab.org}{ryoshida@jlab.org})
\end{abstract}

\maketitle

\section{Introduction}
\label{intro}
Atomic nuclei lie at the core of everything we can see; and at the first level of approximation, their atomic weights are simply the sum of the masses of all the neutrons and protons (nucleons) they contain. Each nucleon has a mass $m_N \sim 1\,$GeV, \emph{i.e}.\ approximately 2000-times the electron mass. The Higgs boson produces the latter, but what produces the masses of the neutron and proton? This is the crux: the vast majority of the mass of a nucleon is lodged with the energy needed to hold quarks together inside it; and that is supposed to be explained by QCD, the strong-interaction piece within the Standard Model.

QCD is unique. It is a fundamental theory with the capacity to sustain massless elementary degrees-of-freedom, \emph{viz}.\ gluons and quarks; yet gluons and quarks are predicted to acquire mass dynamically \cite{Cornwall:1981zr, Aguilar:2015bud, Horn:2016rip}, and nucleons and almost all other hadrons likewise, so that the only massless systems in QCD are its composite NG bosons \cite{Nambu:1960tm, Goldstone:1961eq}, \emph{e.g}.\ pions and kaons.
%Responsible for binding systems as diverse as atomic nuclei and neutron stars, the energy needed to hold the gluons and quarks within these NG modes is not readily apparent.
Responsible for binding systems as diverse as atomic nuclei and neutron stars, the energy associated with the gluons and quarks within these Nambu-Goldstone (NG) modes is not readily apparent.
This is in sharp and fascinating contrast with all other ``everyday'' hadronic bound states, \emph{viz}.\  systems constituted from \emph{up}$\,= u$, \emph{down}$\,=d$, and/or \emph{strange}$\,=s$ quarks, which possess nuclear-size masses far in excess of anything that can directly be tied to the Higgs boson.\footnote{The Higgs mechanism has many attendant puzzles.  For instance, within the Standard Model, nothing constrains the size of the current-quark masses.  The Higgs couplings that produce them are free parameters, determined only after comparisons with data.  The $u$-quark is light; but the $b$-quark is heavy, much heavier than the nucleon.  \emph{A priori}, there is no obvious reason for such disparity.}
%[one question if the Higgs mechanism has a mass size bound theoretically, if this can be explained more or quantified].

In attempting to match QCD with Nature, it is necessary to confront the innumerable complexities of strong, nonlinear dynamics in relativistic quantum field theory, \emph{e.g}.\ the loss of particle number conservation, the frame and scale dependence of the explanations and interpretations of observable processes, and the evolving character of the relevant degrees-of-freedom. Electroweak theory and phenomena are essentially perturbative and thus possess little of this complexity. Science has never before encountered an interaction such as that at work in QCD. Charting this interaction, explaining and understanding everything of which it is capable, can potentially change the way we look at the Universe.

In QCD, the interaction is everything.  Yet, the Lagrangian is remarkably simple in appearance:
\begin{subequations}
\label{QCDdefine}
\begin{align}
{\mathpzc L}_{\rm QCD} & = \sum_{j=u,d,s,\ldots}
\bar{q}_j [i\gamma^\mu D_{\mu} - m_j] q_j - \tfrac{1}{4} G^a_{\mu\nu} G^{a\mu\nu},\\
D_{\mu} & = \partial_\mu + i g \tfrac{1}{2} \lambda^a A^a_\mu\,,\\
\label{gluonSI}
G^a_{\mu\nu} & = \partial_\mu A^a_\nu + \partial_\nu A^a_\mu +
\underline{\textcolor[rgb]{0.00,0.07,1.00}{i g f^{abc}A^b_\mu A^c_\nu}},
\end{align}
\end{subequations}
where $\{q_j\}$ are the quark fields, with $j$ their flavor label and $m_j$ their Higgs-generated current-quark masses. $\{ A_\mu^a, a=1,\ldots, 8 \}$ are the gluon fields, and $\{\tfrac{1}{2} \lambda^a\}$ are the generators of the SU$(3)$ (color/chromo) gauge-group in the fundamental representation. In comparison with quantum electrodynamics (QED), the single, essential difference is the term describing gluon self-interactions, marked as the underlined \underline{\textcolor[rgb]{0.00,0.07,1.00}{blue}} term in Eq.\,\eqref{gluonSI}.
If QCD is correct, as suggested strongly by its ability to describe and predict a wide variety of high-energy phenomena \cite{Tanabashi:2018oca}, for which the theory is perturbative owing to asymptotic freedom, then this term must hold the answers to an enormous number of Nature's basic questions, \emph{e.g}.: what is the origin of visible mass and how is it distributed within atomic nuclei; and what carries the proton's spin and how can the same degrees-of-freedom combine to ensure the pion is spinless? Nowhere are there more basic expressions of \emph{emergence} in Nature.

Treated as a classical theory, chromodynamics is a non-Abelian local gauge field theory. As with all such theories formulated in four spacetime dimensions, no mass-scale exists in the absence of Lagrangian masses for the quarks. There is no dynamics in a scale-invariant theory, only kinematics: the theory looks the same at all length-scales and hence there can be no clumps of anything. Bound states are therefore impossible and, accordingly, our Universe cannot exist.
A spontaneous breaking of symmetry, as realized via the Higgs mechanism, does not solve this problem: the masses of the neutron and proton, the kernels of all visible matter, are roughly 100-times larger than the Higgs-generated current-masses of the light $u$- and $d$-quarks, the main building blocks of protons and neutrons.
%
%[here the question posed on the page before was reiterated: what limits the mass production size of the Higgs mechanism. From our view it may be useful to quantify a bit more why we the pion mass can be mostly Higgs mechanism, for the kaon about half, and small for the nucleon – maybe use the pion mass itself as the scale one expects for the Higgs mass production, with for the kaon the delta(strange – up) quark mass added? Maybe wrong but you can perhaps see the didactical question – although it may be better addressed only cursorily here and more in the next section?].
%

There is a flip-side: the real world's composite NG bosons are (nearly) massless.  Hence, with these systems, the strong interaction's $1\,$GeV mass-scale is effectively hidden. In fact, there is a particular circumstance in which the pseudoscalar mesons $\pi$, $K$, $\eta$ are exactly massless, \emph{i.e}.\ the chiral limit, when the Higgs-generated masses in Eq.\,\eqref{QCDdefine} are omitted.  In this case, perturbative QCD predicts that strong interactions cannot distinguish between quarks with negative or positive helicity. Such chiral symmetry entails an enormous array of consequences, \emph{e.g}.\ the pion would be partnered with a scalar meson of equal mass. However, no state of this type is observed; and, indeed, none of the consequences of this chiral symmetry are found in Nature.  Instead, the symmetry is broken by interactions in QCD. Dynamical chiral symmetry breaking (DCSB) is the agent behind both the massless quarks in QCD’s Lagrangian acquiring a large effective mass \cite{Bhagwat:2003vw, Bowman:2005vx, Bhagwat:2006tu} and the interaction energy between those quarks cancelling their masses exactly so that the composite pion is massless in the chiral limit \cite{Maris:1997hd, Qin:2014vya, Binosi:2016rxz}.

%	Moreover, the pion mass is not mainly Higgs.  The current-mass of the u-quark in the pion is the same as it is in the kaon and the same as it is in the nucleon.
%	The pion mass is zero without a Higgs mechanism; but the tiny Higgs-generated current-mass of the u-quark receives a huge enhancement factor driven by dynamical chiral symmetry breaking.  Note … pi+ = mu + md = just 6 MeV cf. 140 MeV empirically.  So the Higgs-only component of the pion mass is still very small.
%	The scale of the Higgs mechanism for light quarks = 1 MeV.  The scale for the s-quark = 100 MeV.
%	The scale of DCSB = mproton/3.
%	The secret with the pion is that it would normally have a mass of 2*mproton/3 = mass_rho, but most of that mass is canceled by gluon binding effects because the Higgs mechanism is (empirically) weak for light-quarks.

Reinstating the Higgs mechanism, then DCSB is responsible for, \emph{inter alia}: the physical size of the pion mass ($m_\pi \approx 0.15 \,m_N$); the large mass-splitting between the pion and its valence-quark spin-flip partner, the $\rho$-meson ($m_\rho > 5 \,m_\pi$); and the neutron and proton possessing masses $m_N \approx 1\,$GeV. Interesting things happen to the kaon, too. Like a pion, but with one of the light quarks replaced by a $s$-quark, the kaon comes to possess a mass $m_K \approx 0.5\,$GeV. Here a competition is taking place, between dynamical and Higgs-driven mass-generation.

Expanding upon these observations, it is worth highlighting that the physical size of $m_\pi$ is actually far larger than that associated with the Higgs mechanism in the light-quark sector.  Empirically, the scale of the Higgs effect for light quarks is $\sim 1\,$MeV \cite{Tanabashi:2018oca}.  (It is $\sim 100\,$MeV for the $s$-quark.)  As remarked above, $m_\pi = 0$ without a Higgs mechanism; but the current-masses of the light quarks in the pion are the same as they are in the kaons and as in nucleons.  Hence, the simple Higgs-mechanism result is $m_\pi \approx (m_u + m_d)$, yielding a value which is just 5\% of the physical mass.  The physical pion mass emerges as the result of a huge enhancement factor produced by DCSB, which multiplies the current-quark mass contribution.   (See Eq.\,\eqref{GMOR} below and the associated discussion.)  However, the scale of DCSB is $\sim m_N/3$, \emph{i.e}.\ the size of a typical $u$ or $d$ constituent-quark mass; and the special NG-character of the pion means that although it \emph{should} have a mass $\sim (2/3) m_N \approx m_\rho$,  most of that mass is cancelled by gluon binding effects owing to symmetry constraints imposed by DCSB \cite{Roberts:2016vyn}.

These phenomena and features, their origins and corollaries, entail that the question of how did the Universe evolve is inseparable from the questions of how does the $m_N \approx 1\,$GeV mass-scale that characterizes atomic nuclei appear; why does it have the observed value; and, enigmatically, why does the dynamical generation of $m_N$ have seemingly no effect on the composite NG bosons in QCD, \emph{i.e}.\ whence the near-absence of the pion mass?  Theory provides answers but the mechanisms must be confirmed empirically.

In addressing the issues identified above, four central questions arise:
\begin{itemize}
\item How do hadron masses and radii emerge for light-quark systems from QCD?
\item What is the origin and role of dynamical chiral symmetry breaking (DCSB)?
\item What is the interplay of the strong and Higgs-driven mass generation mechanisms?
\item What are the basic mechanisms that determine the distribution of mass, momentum, charge, spin, \emph{etc}. within hadrons?
\end{itemize}

In Sect.\,\ref{Budgets} we provide further analysis of the mass budget of the pion and proton in QCD, and also explain the need for a coherent effort in QCD phenomenology and continuum calculations, in exa-scale computing, as provided by lattice QCD (lQCD), and in experiments. Section~3 compares capabilities foreseen at the EIC with those of the only previous electron-proton collider, \emph{i.e}.\ HERA.  In Sect.\,4, we describe five key experimental efforts aimed at delivering fundamental insights that will reveal answers to the central questions highlighted above:
\begin{itemize}
\item Hadron masses in light quark systems – Measurement: pion and kaon parton distribution functions (PDFs) and pion generalized parton distributions (GPDs);
\item Gluon (binding) energy in NG modes – Measurement: open charm production from pion and kaon;
\item Mass acquisition from DCSB – Measurement: pion and kaon form factors;
\item Strong versus Higgs-driven mass generating mechanisms – Mea\-sure\-ment: valence quark distributions in pion and kaon at large momentum fraction $x$;
\item Timelike analog of mass acquisition – Measurement: fragmentation of a quark into pions or kaons.
\end{itemize}
Finally, in Sect.\,5, we provide a summary of the ideas and opportunities described herein.

\section{Mass Budgets}
\label{Budgets}
In field theory, scale invariance is expressed in conservation of the dilation current\footnote{As typical in discussions of strong QCD, hereafter we employ standard Euclidean metric conventions, \emph{e.g}.: for Dirac matrices, $\{\gamma_\mu,\gamma_\nu\} = 2 \delta_{\mu\nu}$, $\gamma_\mu^\dagger = \gamma_\mu$; and $a\cdot b = \sum_{i=1}^4 a_i b_i$.  A timelike vector, $p_\mu$, has $p^2<0$.}
\begin{equation}
\partial_\mu D_\mu = \partial_\mu (T_{\mu\nu} x_\nu)  = T_{\mu\mu} = 0\,,
\end{equation}
where $T_{\mu\nu}$ is the theory’s energy-momentum tensor, which satisfies $\partial_\mu T_{\mu\nu}=0$ owing to energy and momentum conservation.  The catastrophic consequences of scale invariance -- \emph{e.g}., bound states being impossible -- are avoided in Nature through the agency of quantum effects. %
In quantizing QCD, regularization and renormalization of (ultraviolet) divergences introduces a mass-scale, $\zeta$.  Consequently, mass-dimensionless quantities and other ``constants'' become dependent on $\zeta$. This is ``dimensional transmutation''.  It entails the appearance of a \emph{trace anomaly}:
\begin{equation}
 T_{\mu\mu} = \beta(\alpha(\zeta)) \, \tfrac{1}{4}\, G_{\alpha\beta}^a G_{\alpha\beta}^a =: \Theta_0 \,,
\end{equation}
where $\beta(\alpha(\zeta))$ is QCD's $\beta$-function and $\alpha(\zeta)$ is the associated running-coupling, which indicates that \underline{a mass-scale is born}.

This mass-scale is exhibited in the gauge-boson vacuum polarization. In QED, the photon vacuum polarization does not possess an infrared mass-scale, and dimensional transmutation serves merely to produce the very slow running of the QED coupling, \emph{i.e}.\ any dynamical violation of the conformal features of QED is very small and hence the trace anomaly is negligible.  In contrast, owing to gauge sector dynamics, a Schwinger mechanism is active in QCD \cite{Cornwall:1981zr, Aguilar:2015bud}, so that the QCD trace anomaly expresses a mass-scale which is, empirically, very significant. The relation between QCD's gluons (gauge-bosons) and the trace anomaly could be elucidated by experimental and theoretical studies of hadron states in which the presence of glue determines the quantum numbers, such as hybrid mesons and baryons.

Owing to Einstein's energy-mass relation: $E=mc^2$, one might consider a rest-frame decomposition of the proton's mass into contributions from various components of $T_{\mu\nu}$ \cite{Ji:1995sv}; and a contemporary lQCD calculation reports \cite{Yang:2018nqn}:
trace anomaly = 23(1)\%;
quark current-mass term = 9(2)\%;
gluon kinetic + potential energy = 36(6)\%;
and quark kinetic + potential energy = 32(6)\%.
%%
%In one such approach, the following contributions have been estimated \cite{Ji:1995sv}:  trace anomaly = 20\%; quark current-mass term = 17\%; gluon kinetic + potential energy = 34\%; and quark kinetic + potential energy = 29\%.
%
%\footnote{A contemporary lQCD calculation reports \cite{Yang:2018nqn}: trace anomaly = 23(1)\%, quark current-mass term = 9(2)\%; gluon kinetic + potential energy = 36(6)\%; and quark kinetic + potential energy = 32(6)\%.}
%
This decompostion, however, should be interpreted with care because (\emph{i}) they depend on the reference frame and the choice of renormalization scale and (\emph{ii}) the gluons in the trace anomaly and in the kinetic and potential energy are seemingly being treated as separate entities \cite{Lorce:2017xzd}.   Furthermore, it can equally be argued that in the chiral limit, using a parton model basis \cite{Roberts:2016vyn}: \emph{the entirety of the proton mass is produced by gluons and due to the trace anomaly};
\begin{equation}
\label{protonmassanomaly}
  \langle P(p)| \Theta_0 | P(p)\rangle = - p_\mu p_\mu = m_N^2\,.
\end{equation}
Crucially, it might be possible to access the trace anomaly contribution to the proton mass through the production of $J/\Psi$ and $\Upsilon$ mesons at threshold \cite{Kharzeev:1998bz, Joosten:2018gyo}, through which color van der Waals forces could be accessible \cite{TarrusCastella:2018php}.

Given that the pion is massless in the chiral limit, then a rest-frame decomposition of the contributions to its mass is impossible: massless particles do not have a rest frame.  One might attempt this for the physical-mass pion, \emph{i.e}.\ with the Higgs mechanism turned-on, but then it is crucial to recall the Gell-Mann--Oakes--Renner relation \cite{GellMann:1968rz}, which in modern form is
\begin{equation}
\label{GMOR}
m_\pi^2 = (m_u + m_d) \frac{-\langle \bar q q \rangle}{f_\pi^2}\,,
\end{equation}
where $f_\pi$ is the pion's leptonic decay constant and $\langle \bar q q \rangle$ is the chiral condensate \cite{Brodsky:2012ku}.

Equation\,\eqref{GMOR} entails that \emph{all} the pion's mass-squared is generated by the Higgs-connected mass term in QCD’s Lagrangian.  It is impossible, therefore, that only half (or any other fraction) of $m_\pi$ is generated by this same term \cite{Yang:2014xsa}.  Evidently, a rest-frame decomposition of the pion's mass is a theoretical challenge and can be open to misinterpretation. Notwithstanding this, there are crucial open questions.  For instance, both $f_\pi$ and $\langle \bar q q \rangle$  are order parameters for DCSB, providing a magnification factor for the Higgs-generated current-quark masses, which is empirically known to be very large.  So, what is the relationship between the size of this magnification factor and the gluon distribution in the pion?  In principle, such information could be obtained using deep inelastic scattering off pions (and kaons) to constrain the associated generalized parton distributions, whose leading moments can be related to expectation values of QCD's energy momentum tensor in the target hadron \cite{Lorce:2015lna}.

Plainly, further guidance is required before one can properly unfold the trace anomaly's contribution to the masses of the proton, pion and kaon. This is highlighted by the following series of observations. Consider the chiral limit, in which case $m_\pi=0$, so that
\begin{equation}
\label{pionmassanomaly}
  \langle \pi(k)| \Theta_0 | \pi(k)\rangle = - k_\mu k_\mu = m_\pi^2 =0 \,,
\end{equation}
\emph{i.e}.\ the expectation value of the trace anomaly in the chiral-limit pion is identically zero.  One na\"ive interpretation of Eq.\,\eqref{pionmassanomaly} is that in the chiral limit the gluons disappear and thus contribute nothing to the pion mass. However, in the presence of a trace anomaly, both gluons and quarks acquire masses dynamically, so this is unlikely.
Nonetheless, were it true, then one would be faced with some interesting conundrums, \emph{e.g}.\ it would mean that at large renormalization scales, $\zeta \gg m_N$, the proton is full of gluons, whereas the pion at such large scales is empty of gluons, and remains so despite the fact that QCD evolution \cite{Dokshitzer:1977, Gribov:1972, Lipatov:1974qm, Altarelli:1977} suggests gluons dominate within every hadron on the neighborhood $m_N/\zeta \simeq 0$ \cite{Altarelli:1981ax}. Such consequences could be tested experimentally: are there gluons in the pion or not?  Herein, we will outline a set of five key measurements that can address such basic issues.

A more likely explanation, and arguably the correct one, is that the expectation value of the scale anomaly vanishes in the chiral-limit pion owing to cancellations between competing effects associated with different interaction mechanisms, which are exact in the pion channel because of DCSB.
Indeed, existing pion structure data from Drell-Yan processes, albeit limited, have been analyzed to separate valence quark, sea quark and gluon parton distributions.  It indicates that about 40\% of the momentum of the pion is carried by gluons at a scale $Q^2 = 5\,{\rm GeV}^2$ \cite{Sutton:1991ay, Barry:2018ort}, roughly the same as their contribution in a proton -- a finite and large gluon contribution.

In either event, with Eqs.\,\eqref{protonmassanomaly} and \eqref{pionmassanomaly} one is confronted with a peculiar dichotomy, which insists that no answer to the question ``Whence the proton's mass?'' is complete unless it simultaneously solves the additional puzzle ``Whence the \emph{absence} of a pion mass?'' The natural nuclear-physics mass-scale, $m_N$, must emerge simultaneously with apparent preservation of scale invariance in an intimately related system such as the (chiral limit) pion and kaon. Furthermore, in the chiral limit, the aforementioned cancellations must occur in the pion irrespective of the size of this mass scale, $m_N$.

These statements hold with equal force on a sizeable neighborhood of the chiral limit because hadron masses are continuous functions of the current-quark masses. Then, using the Gell-Mann--Oakes--Renner relation \cite{GellMann:1968rz}, as done above, it follows that the pion, or any other NG boson, has the peculiar property that the entirety of its mass owes to the current-quark mass term in QCD’s Lagrangian, Eq.\,\eqref{QCDdefine}. It is natural to compare this result with that for the pion's valence-quark spin-flip partner, the $\rho$-meson, for which just 6\% of its mass-squared is directly tied to the current-quark mass term \cite{Flambaum:2005kc}.

The key to understanding Eq.\,\eqref{pionmassanomaly} is a quartet of Goldberger-Treiman-like (GT) relations \cite{Maris:1997hd, Qin:2014vya}, the best known of which states:
\begin{equation}
\label{EqDCSB}
m \simeq 0 \; \left| \; f_\pi E_\pi(\ell;0) = B(\ell^2)
\right. ,
\end{equation}
where $E_\pi$ is the leading piece of the pion's Bethe-Salpeter amplitude, and $B$ is the scalar piece of the dressed-quark self-energy. This equation is exact in chiral QCD and expresses the fact that Goldstone's theorem is fundamentally an expression of equivalence between the quark one-body problem and the two-body bound-state problem in QCD's color-singlet flavor-nonsinglet pseudoscalar channel. An amazing consequence is that the properties of the nearly-massless pion are the cleanest expression of the mechanism that is responsible for almost all the visible mass in the Universe. It is notable that a rudimentary form of this equation can be found in the work which brought Nambu one half of the 2008 Nobel Prize in Physics for ``the discovery of the mechanism of spontaneous broken symmetry in subatomic physics'' \cite{Nambu:1961tp}.

With the quartet of GT relations in hand, one can construct an algebraic proof \cite{Maris:1997hd, Binosi:2016rxz}, that at any and each order in a symmetry-preserving truncation of those equations in quantum field theory necessary to describe a pseudoscalar bound state, there is a precise cancellation between the mass-generating effect of dressing the valence-quarks which constitute the system and the attraction generated by the interactions between them. This guarantees the ``disappearance'' of the scale anomaly in the pion in the chiral limit through cancellations between one-body dressing and two-body interaction effects that sum precisely to zero because chiral symmetry is dynamically broken -- the dressed quark and anti-quark masses in the pion are canceled by a (negative) binding energy. An analogy with quantum mechanics emerges: the mass of a QCD bound-state is the sum of the mass-scales characteristic of the constituents plus some (negative and sometimes large) binding energy.

\begin{figure}[t]
\vspace*{3ex}

\centerline{%
\includegraphics[clip, width=0.44\textwidth]{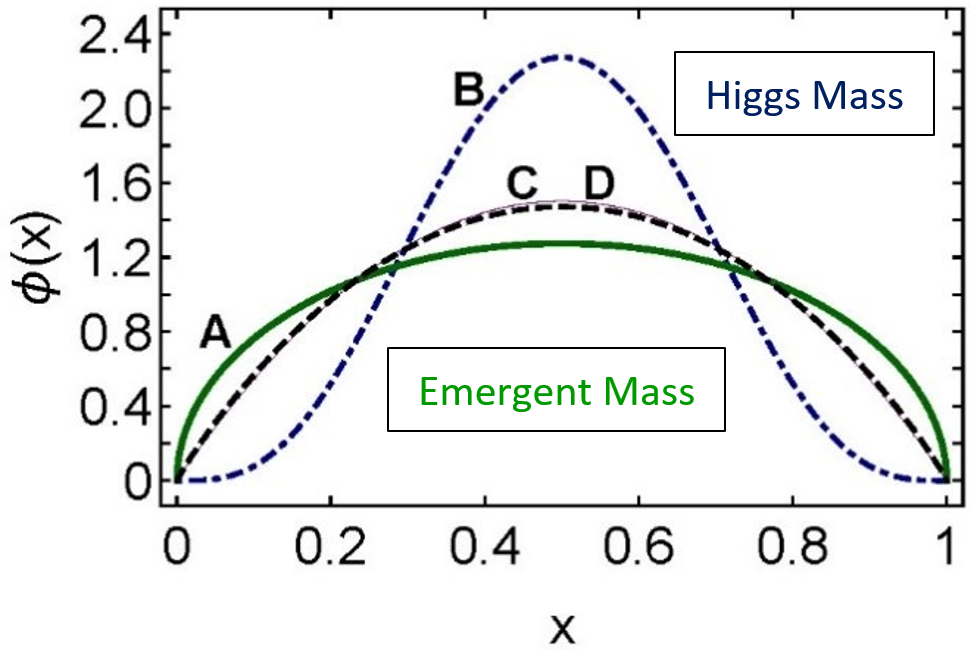}}
\caption{\label{figPDAs}
Twist-two parton distribution amplitudes at a resolving scale $\zeta=2 \,$GeV$=:\zeta_2$. \textbf{A} solid (green) curve – pion $\Leftarrow$ emergent mass generation is dominant; \textbf{B} dot-dashed (blue) curve – $\eta_c$ meson $\Leftarrow$ Higgs mechanism is the primary source of mass generation;  \textbf{C} solid (thin, purple) curve -- asymptotic profile, 6x(1 - x); and \textbf{D} dashed (black) curve – ``heavy-pion'', \emph{i.e}.\ a pion-like pseudo-scalar meson in which the valence-quark current masses take values corresponding to a strange quark $\Leftarrow$ the boundary, where emergent and Higgs-driven mass generation are equally important.
}
\end{figure}

Since QCD's interactions are universal and the same in all hadrons, similar cancellations must take place within the proton. However, in the proton channel there is no symmetry that requires the cancellations to be complete. Hence, the proton's mass has a value that is typical of the magnitude of scale breaking in QCD's one body sectors, \emph{viz}.\ the dressed-gluon and -quark mass scales. The picture described here may be called the ``DCSB paradigm''.  It provides a basis for understanding why the mass-scale for strong interactions is vastly different to that of electromagnetism, why the proton mass expresses that scale, and why the pion is nevertheless unnaturally light.

In this picture, no significant mass-scale is possible in QCD unless one of commensurate size is expressed in the dressed-propagators of gluons and quarks. It follows that the mechanism(s) responsible for the generation of mass in QCD can be exposed by measurements that are sensitive to such dressing effects.

This potential is offered by many observables, including hadron elastic and transition form factors; but as an illustrative example, consider a particular class of meson ``wave functions'', \emph{i.e}.\ twist-two parton distribution amplitudes (PDAs), a number of which are depicted in Fig.\,\ref{figPDAs}. This image answers the following question: When does the Higgs mechanism begin to influence mass generation? In the limit of infinitely-heavy quark masses; namely, when the Higgs mechanism has overwhelmed every other mass generating force, the PDA becomes a $\delta$-function at $x = \tfrac{1}{2}$. The sufficiently heavy $\eta_c$ meson, constituted from a valence charm-quark and its antimatter partner, feels the Higgs mechanism strongly. On the other hand, contemporary continuum- and lattice-QCD calculations predict that the PDA for the light-quark pion is a broad, concave function \cite{Brodsky:2006uqa, Chang:2013pq, Zhang:2017bzy, Jia:2018ary, Bali:2018spj}. Such features are a definitive signal that pion properties express emergent mass generation. The remaining example in Fig.\,\ref{figPDAs} shows that the PDA for a system composed of $s$-quarks almost matches that of QCD’s asymptotic (scale-free) limit: this system lies at the boundary, with strong (emergent) mass generation and the weak (Higgs-driven) mass playing a roughly equal role.
%Mikhailov:1986be, Petrov:1998kg, Brodsky:2006uqa, Chang:2013pq

\begin{figure}[t]
\centerline{%
\includegraphics[clip, width=0.45\textwidth]{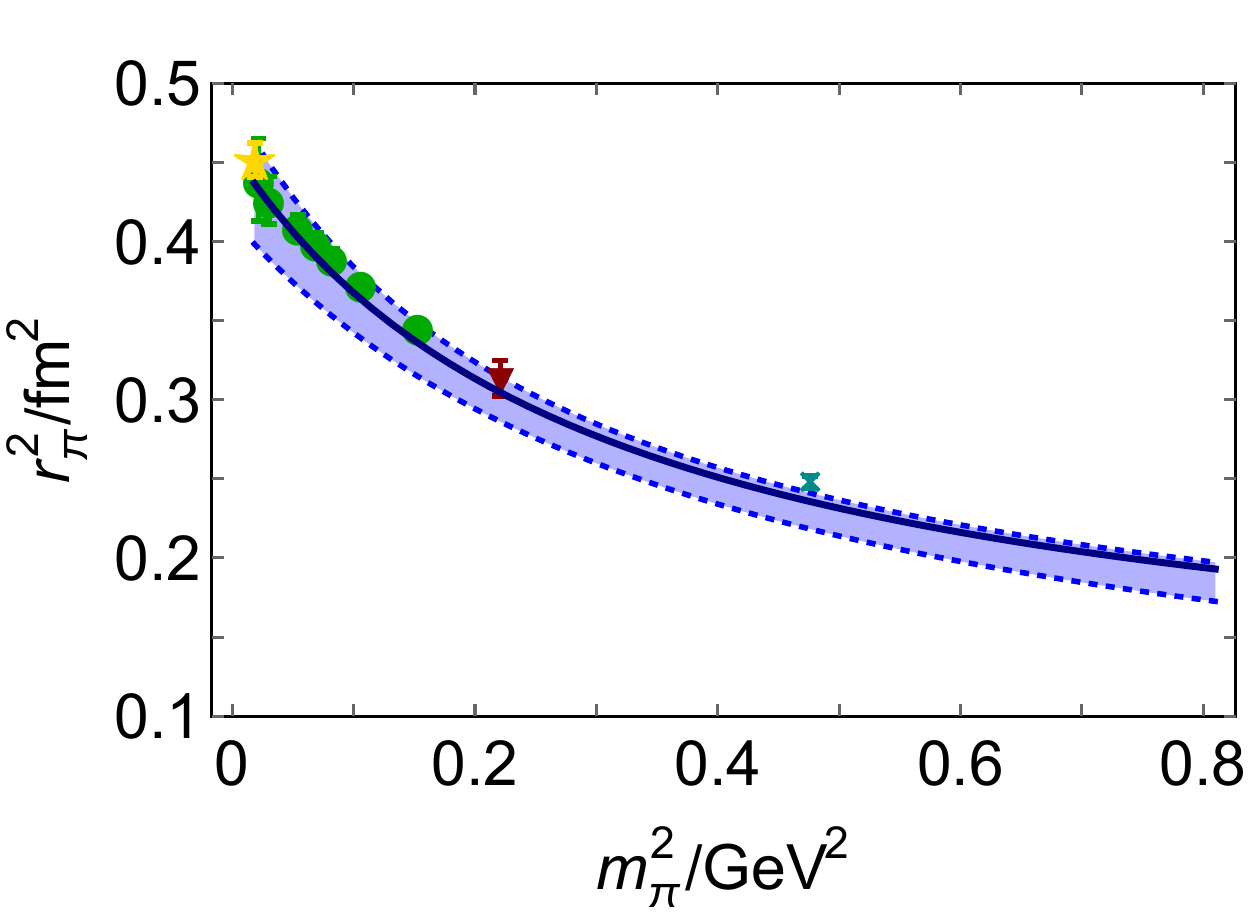}}
\caption{\label{figradius}
Lattice-QCD computations of the pion’s electromagnetic charge radius (green circles \cite{Wang:2018pii},
%blue up-triangle \cite{ChakrabortyPrivate},
red down-triangle \cite{Chambers:2017tuf}, cyan cross \cite{Koponen:2017fvm}) as a function of $m_\pi^2$, compared with a continuum theory prediction \cite{Chen:2018rwz} (blue curve within bands, which indicate response to reasonable parameter variation).  The continuum analysis establishes $f_\pi r_\pi \approx\,$constant, from which it follows that the size of a Nambu-Goldstone mode decreases in inverse proportion to the active strength of the dominant mass generating mechanism.  The empirical value of $r_\pi$ is marked by the gold star.
}
\end{figure}

These observations indicate that comparisons between distributions of truly light quarks and those describing strange quarks are ideally suited to exposing measurable signals of dynamical mass generation.

In selecting measurements that will enable the origin of mass in the pion and kaon to be identified and the mass distributions charted, one will also be led naturally to experiments and analyses that reveal the distribution of charge, momentum, spin, \emph{etc}., within these most fundamental of bosons.  For example, measuring the pion’s electromagnetic size and mapping its charge distribution have long been central problems in nuclear physics.  The radius is known \cite{Tanabashi:2018oca}, banner experiments at the Thomas Jefferson National Accelerator Facility (JLab) have provided precise data on the elastic electromagnetic form factor out to $Q^2 \approx 2.5\,{\rm GeV}^2$ \cite{Volmer:2000ek, Horn:2006tm, Horn:2007ug, Blok:2008jy, Huber:2008id}, planned experiments will extend this upper bound to $Q^2 \approx 8.5\,{\rm GeV}^2$ \cite{E12-06-101, E12-07-105, Horn:2017csb}, and continuum- and lattice-QCD analyses are making predictions that connect these properties to the origin of mass, through the pion's leptonic decay constant and the distribution amplitudes in Fig.\,\ref{figPDAs}.  Recent progress is illustrated in Fig.\,\ref{figradius}, which displays contemporary lQCD \cite{Wang:2018pii, Chambers:2017tuf, Koponen:2017fvm}  and continuum  computations \cite{Chen:2018rwz} of the pion's charge radius, and correlates them with the source of mass in the Standard Model.  The program described herein will therefore have a wide-ranging impact on our understanding of the strong forces that shape hadrons, nuclei and nuclear matter.

\begin{figure}[t]
\centerline{%
\includegraphics[clip, width=0.30\textwidth]{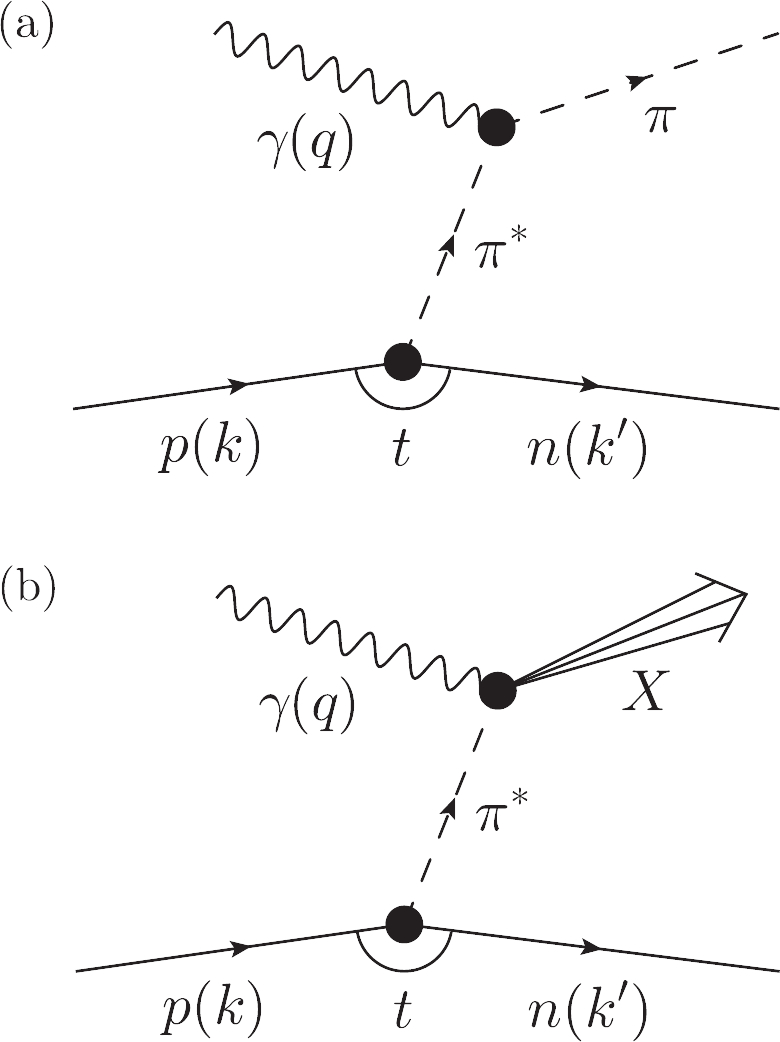}}
\caption{\label{figSullivan}
Sullivan processes.  In these examples, a nucleon's pion cloud is used to provide access to the pion's (a) elastic form factor and (b) parton distribution functions.  $t = –(k-k^\prime)^2$ is a Mandelstam variable and the intermediate pion, $\pi^\ast(P=k-k^\prime)$, $P^2= –t$, is off-shell.
}
\end{figure}

\section{Capacity of EIC and Comparison with HERA }
\subsection{Pion and Kaon Sullivan Process}
\label{secPiKSullivan}
In specific kinematic regions, the observation of recoil nucleons (hyperons) in the semi-inclusive reaction $e p \to e^\prime (n\,{\rm or}\,Y) X$ can reveal features associated with correlated quark-antiquark pairs in the nucleon, referred to as the ``nucleon’s meson cloud'', or the ``five-quark component of the nucleon wave function''.  In particular, according to current models, at low values of the four-momentum, $t$, transferred from the initial proton, $p$, to the final neutron, $n$, or hyperon, $Y$, the cross-section displays behavior characteristic of meson (pion \emph{vs}.\ kaon, respectively) pole dominance. The electron deep-inelastic-scattering (DIS) off the meson cloud of a nucleon target is called the Sullivan process \cite{Sullivan:1971kd}.  Illustrated in Fig.\,\ref{figSullivan}, it is typically interpreted such that the nucleon parton distributions necessarily contain a mesonic parton content \cite{Melnitchouk:1995en, Salamu:2014pka, Salamu:2018cny}. To access the pion or kaon partonic content via such a structure function measurement requires scattering from a meson target.

\begin{figure}[t]
\centerline{%
\includegraphics[clip, width=0.44\textwidth]{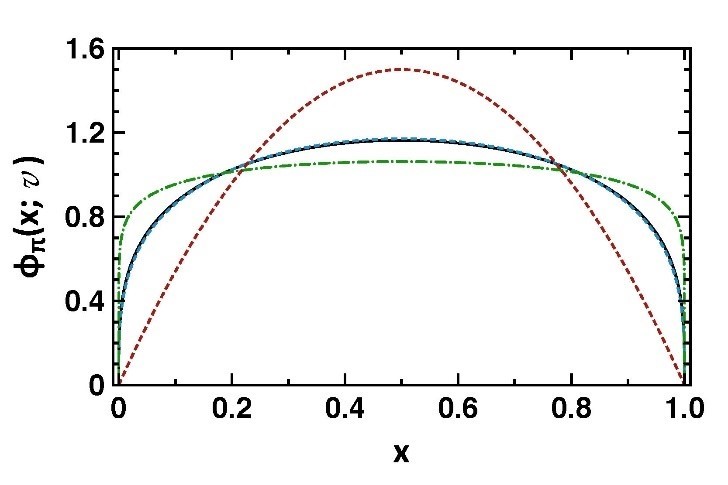}}
\caption{\label{figPDAvirtuality}
Virtuality-dependence of pion twist-two PDA.  Solid (blue) curve: $v_\pi =0$ result; and dot-dashed (green) curve, PDA at $v_\pi=31$. Even this appreciable virtuality only introduces a modest rms relative-difference between the computed PDAs; namely, 13\%.  Measured equivalently, the zero virtuality result differs by 34\% from that appropriate to QCD's asymptotic limit (dotted, red curve).
}
\end{figure}

Theoretically, the Sullivan process can provide reliable access to a meson target as $t$ becomes space-like, if the pole associated with the ground-state meson remains the dominant feature of the process and the structure of the related correlation evolves slowly and smoothly with virtuality. To check whether these conditions are satisfied empirically, one can take data covering a range in $t$, particularly low $|t|$, and compare with phenomenological and theoretical expectations. Theoretically, a recent calculation \cite{Qin:2017lcd} explored the circumstances under which these conditions should be satisfied.  Defining pion virtuality as $v_\pi = (m_\pi^2-t)/m_\pi^2$, it was found that, for $v_\pi \leq 30$, which corresponds to $-t \leq 0.6\,$GeV$^2$, all changes in pion structure are modest so that a well-constrained experimental analysis should be reliable (see Fig.\,\ref{figPDAvirtuality}). Similar analyses for the kaon indicate that Sullivan processes can provide a valid kaon target for $-t \leq 0.9\,$GeV$^2$.

Experimentally, one needs to ensure that the Sullivan process is a valid tool for meson structure experiments.  For this to be true in elastic form factor measurements, Fig.\,\ref{figSullivan}(a), one must ensure that the virtual photon is longitudinally polarized.  At the high $Q^2$, $W$ accessible with the EIC, phenomenological models predict $\sigma_L \gg  \sigma_T$ at small $-t$.  A practical method of isolating the longitudinal virtual photon is to use a model to distinguish the dominant differential cross-section, $d \sigma_L/dt$, from the measured, unseparated differential cross-section, $d \sigma/dt$.  Focusing on the pion because the kaon is similar, one can then experimentally validate the model, \emph{i.e}.\ the condition $\sigma_L \gg  \sigma_T$, by using the $\pi^-/\pi^+$ ratio of charged pion data extracted from electron-deuteron beam collisions in the same kinematics as charged pion data from electron-proton collisions.  $G$-parity conservation entails that the $\pi^-/\pi^+$ ratio will differ from unity if $\sigma_T$ is large or if there are significant non-pole contributions.  Data on $Q^2\in [1,10]\,$GeV$^2$ from exclusive pion (and kaon) experiments at the 12\,GeV JLab can be used to provide additional assistance in validating the model.  Further details are provided elsewhere \cite[Sec.\,2]{Horn:2016rip}.
On the other hand, with structure functions, Fig.\,\ref{figSullivan}(b), one can work with the entire differential cross-section, which is transverse in the Bjorken limit, and depend upon both the phenomenology and theory which predict that meson structure can reliably be extracted on a sizeable low-$|t|$ domain and comparisons with results from other experimental techniques on their common domain.

%The most practical choice to isolate the longitudinal virtual photon is to use a model to isolate the dominant differential cross section dL/dt from the measured unseparated d/dt. The same principle was used for the Cornell experiments (C.J. Bebek et al. Phys. Rev. D17 (1978) 1693). One can validate the model, i.e., the condition L >> T, experimentally using the ratio of charged-pion -/+ data extracted from electron-deuteron beam collisions in the same kinematics as charged pion data from electron-proton beam collisions. By G-parity conservation the -/+ ratio will be diluted from unity if T is not small, or if there are significant non-pole contributions. Data in the region of Q2=1-10 GeV2 from 12 GeV exclusive pion and kaon experiments can provide additional information to validate the model.

\subsection{EIC Detection Capabilities}
\label{secEICdetection}
In the case of a proton to neutron Sullivan process, used to tag a virtual pion target, the final state neutron moves forward with a large part of the initial beam energy. For an EIC with proton beam energy $E_{\rm b}=100\,$GeV, this means detecting near to 100\,GeV neutrons in a zero-degree calorimeter (ZDC). The ZDC must reconstruct the energy and position well enough to help constrain both the scattering kinematics and the 4-momentum of the pion.

A 35\%$/\surd E_{\rm b}$ energy resolution will constrain the neutron energy to 3.5\%; this resolution for high-energy neutrons is known to be achievable for hadron calorimeters and directly translates into the achievable resolution in $x$ \cite{Andresen:1989qr, DiCapua:1995vc, Lee:2017shn}.  A forward detector is integrated with the interaction region and provides excellent coverage and resolution.\footnote{Like the version designed for the JLab EIC (JLEIC); see also the selected results by C.\,Weiss \emph{et al}., Spectator Tagging Project, \href{https://www.jlab.org/theory/tag/}{https://www.jlab.org/theory/tag/\,.}}

Protons and charged fragments passing through the final focusing quadrupole are analyzed by a magnetic dipole field and detected with longitudinal momentum resolution $\delta p_L/p_L \sim 10^{-4}$, transverse momentum resolution of $\delta p_T \sim 20\,$MeV, and complete coverage down to $p_T=0$. The position resolution needs depend on the positioning of the ZDC; but assuming a placement of the detector at 50\,meters from the interaction point, a resolution of a few mm, which has been demonstrated \cite{Chekanov:2007tv} for forward neutrons at HERA with proton beam energies of 820\,GeV, will achieve $\sim 0.1\,$mrad angular resolution at the vertex.

A nice consistency check can further be provided by tagging two protons following an electron-deuteron beam collision to measure the neutron to proton Sullivan process, a relatively easy detection for an EIC with analysis of two charged particles.

\begin{figure}[t]
\vspace*{2ex}

\centerline{%
\includegraphics[clip, width=0.48\textwidth]{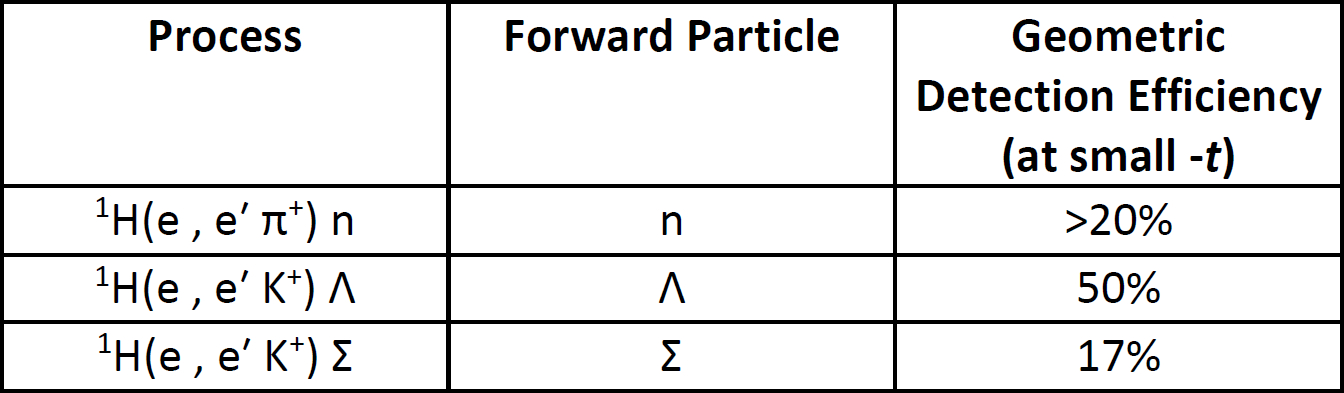}}
\caption{\label{figGeometric}
Geometric acceptances for detection of leading neutrons and the decay products of $\Lambda$ and $\Sigma$ particles in the integrated JLEIC detector concept, to tag the pion and kaon Sullivan processes.
}
\end{figure}

For the $p\to \Lambda$ Sullivan process, as required to tag the kaon cloud, the decay products of the $\Lambda$ must be tracked through the very forward spectrometer. Some initial acceptance studies were done that indicate a good geometric acceptance for the decay products (see Fig.\,\ref{figGeometric}), but further studies need to be completed in order to quantify the performance in terms of particle identification and resolution of the current EIC detectors. Note that the various processes to tag the pion and kaon can be measured simultaneously, and that the geometric acceptances are never fully 100\% as a forward-going neutron or a decay product can end up in the bore of the various interaction magnets.

\subsection{Anticipated Statistical Precision of Pion and Kaon Structure Function Results}
The projected brightness for a high-luminosity EIC is nearly three orders-of-magnitude above that of HERA, $10^{34}\,$e-nucleons/cm$^2$/s versus $10^{31}\,$e-nucleons/cm$^2$/s. In addition, detection fractions for the pion and kaon Sullivan-process products are constructed to be better at EIC as compared to HERA. For instance, the anticipated leading neutron detection is well beyond the 20\% leading neutron detection efficiency of HERA, and the geometric detection efficiency for the kaon Sullivan-process products is also roughly 20\% or better at EIC (see Fig.\,\ref{figGeometric}). Notably, the detection of two-charged protons following $e-d$ collisions to map the (negatively-charged) pion Sullivan-process is nearly 100\% efficient.

\begin{figure}[t]
\centerline{%
\includegraphics[clip, width=0.44\textwidth]{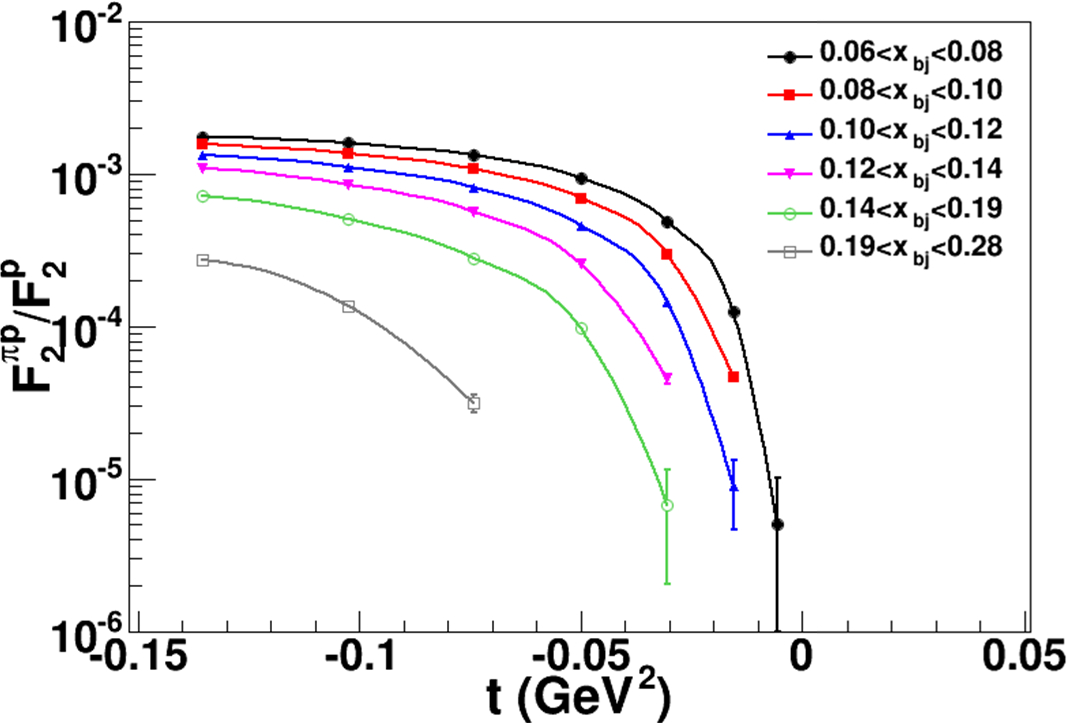}}
\caption{\label{figF2x}
Ratio of the component of the $F_2$ structure function related to the pion Sullivan process as compared to the proton $F_2$ structure function in the low-$t$ vicinity of the pion pole, as a function of $t$ for various values of Bjorken-$x$.
}
\end{figure}

Lastly, the duration of active data taking for ZDCs at HERA was limited. Hence, one anticipates roughly a four orders-of-magnitude additional reach in pion structure function measurements at EIC versus what was achieved in the pioneering measurements at HERA. This directly balances the ratio of tagged pion structure function measurements through the Sullivan process in various bins of $–t$ (with bin size of $0.02\,$GeV$^2$) to $F_2$ proton structure function measurements (see Fig.\,\ref{figF2x}). With a suitable detector configuration, access to high $x_\pi$ ($\to 1$) will be possible, allowing overlap with fixed-target experiments \cite{Keppel:2015, Keppel:2015B, McKenney:2015xis}.  Overlap with Drell-Yan measurements will also settle the unknown pion flux factor associated with the Sullivan process measurements. Kaon structure was not studied at HERA; but the ratio of kaon structure function (under the condition of a $\Lambda$ detection) to the proton structure function at small $-t$ is similar to that for the pion Sullivan process ($\sim 10^{-3}$).   Hence, one would anticipate both pion and kaon structure function measurements as functions of $(–t, x, Q^2)$ at a high-luminosity ($10^{34}$ or more) EIC to be of similar statistical precision as the well-known, textbook HERA proton $F_2$ structure function measurements.  One should therefore be able to constrain the gluon distributions in the pion and kaon.

\section{Key EIC Measurements}
We now outline a science program, consisting of (a series of) five key measurements, whose composition emerged from discussions during a series of workshops focused on studies of pion and kaon structure at the EIC.  These five key measurements were chosen because of their potential to deliver fundamental insights into the origin of the mass of those hadron bound states whose existence is crucial to the evolution of our universe.

\subsection{Pion and Kaon PDFs and Pion GPDs}
%
%Sec.\,\ref{Budgets} explained that Eq.\,\eqref{EqDCSB} is a defining feature of the DCSB paradigm for mass generation, according to which no significant mass-scale is possible in QCD unless one of commensurate size is expressed in the dressed-propagators of gluons and quarks; and, hence, the mechanism(s) responsible for the emergence of mass can be uncovered by experiments that are sensitive to such dressing effects.  The cleanest such measurement is a determination of ${\mathpzc u}^\pi(x;\zeta)$, the pion's valence-quark parton distribution function (PDF).  For example, akin to the PDA effect illustrated in Fig.\,\ref{figPDAs}, significant broadening of ${\mathpzc u}^\pi(x;\zeta)$ with respect to the scale-free profile is a unique expression of emergent mass \cite{Ding:2019lwe}.  Moreover, as Sec.\,\ref{SecKaonStructure} will make clear, the potential of such experiments is greatly enhanced if one includes similar kaon measurements.
%
The mass of the pion is roughly 140\,MeV, of the kaon 493\,MeV, and of the proton 939\,MeV. In the chiral limit, the mass of the proton is entirely given by the trace anomaly in QCD, Eq.\,\eqref{protonmassanomaly}. The mass of the pion has, in this same limit, either no contribution at all from the trace anomaly or, more likely, a cancellation of terms occurs in order to ensure the pion is massless, Eq.\,\eqref{pionmassanomaly}.

Beyond the chiral limit, a decomposition of the proton mass budget has been suggested, expressing contributions from quark and gluon energy and quark masses \cite{Ji:1995sv}. With the various quark (flavor) and gluon distributions in the proton reasonably well known, the largest uncertainty here lies with the trace anomaly contribution.

\begin{figure}[t]
\vspace*{2ex}

\centerline{%
\includegraphics[clip, width=0.44\textwidth]{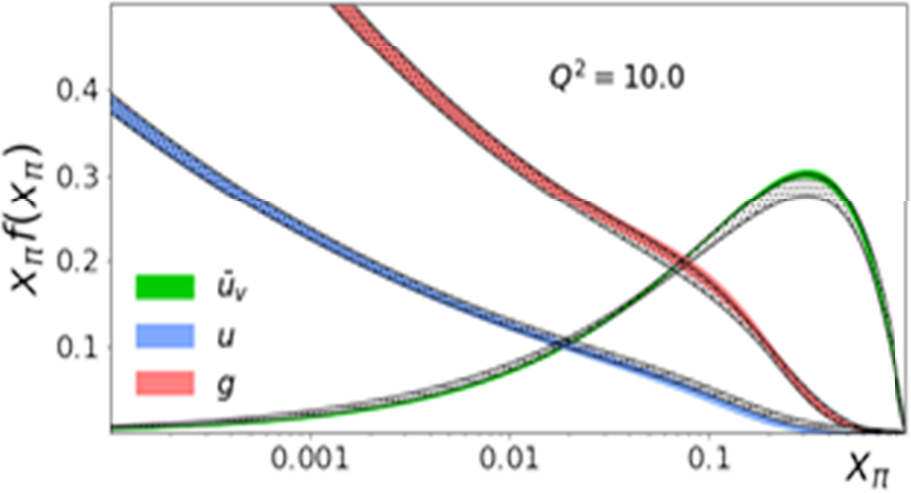}}
\caption{\label{figF7}
A sample EIC extraction of valence quark, sea quark and gluon PDFs in the pion, at a scale $Q^2 =10\,$GeV$^2$. The extraction is done with the following assumptions on PDFs: the $u$ PDF equals the $\bar d$ PDF in the pion and the $\bar u$ PDF is the same as the other sea quark PDFs ($d$, $s$ and $\bar s$). The extraction at $x_\pi  < 10^{-2}$, at this $Q^2$ scale, is constrained by the existing HERA data.
}
\end{figure}

For the pion, further guidance on the magnitude of quark and gluon energy can, as for the proton, be determined from measurements of the pion and kaon structure functions, with resulting constraints on quark and gluon PDFs, over a large range of $x_\pi$ and momentum-transfer squared, $Q^2$. This is accessible for the EIC, roughly covering down to $x_\pi = 10^{-3}$ at $Q^2 = 1\,$GeV$^2$ and up to $x_\pi=1$ at $Q^2 = 1000\,$GeV$^2$.  A sample extraction of valence quark, sea quark and gluon distributions from projected EIC data is given in Fig.\,\ref{figF7}. Data from electron-deuteron collisions, tagged with two high-energy protons, can further give access to the oppositely-charged pion Sullivan process and provide a comparison.

As Sec.\,\ref{Budgets} explained, Eq.\,\eqref{EqDCSB} is a defining feature of the DCSB paradigm for mass generation, according to which no significant mass-scale is possible in QCD unless one of commensurate size is expressed in the dressed-propagators of gluons and quarks; and, hence, the mechanism(s) responsible for the emergence of mass can be uncovered by experiments that are sensitive to such dressing effects.  The cleanest such measurement is a determination of ${\mathpzc u}^\pi(x;\zeta)$, the pion's valence-quark parton distribution function (PDF) at the probe scale $\zeta$.  Akin to the PDA effect illustrated in Fig.\,\ref{figPDAs}, significant broadening of ${\mathpzc u}^\pi(x;\zeta)$ with respect to the scale-free profile and contrasts with the magnitude of such effects in ${\mathpzc u}^K(x;\zeta)$ can serve as unique expressions of emergent mass \cite{Ding:2019lwe}.

%Moreover, as Sec.\,\ref{SecKaonStructure} will make clear, the potential of such experiments is greatly enhanced if one includes similar kaon measurements.

%"Sec. 2 explained that Eq. (7) is a defining feature of the DCSB paradigm for mass generation, according to which no significant mass-scale is possible in QCD unless one of commensurate size is expressed in the dressed-propagators of gluons and quarks; and, hence, the mechanism(s) responsible for the emergence of mass can be uncovered by experiments that are sensitive to such dressing effects. The cleanest such measurement is a determination of $u^\pi(x,\dzeta)$, the pion's valence-quark parton distribution function (PDF) at scale $\dzeta$. Akin to the PDA effect illustrated in Fig. 1, significant broadening of $u^\pi(x,$\dzeta)$ (and also $u^K(x,\dzeta)$) with respect to a scale-free profile can be a unique expression of emergent mass [62]."

Additionally, measurement of the neutral-current parity-violating asymmetry could provide a further consistency check for the pion PDFs and also, for the kaon, enable $u$ and $s$, $\bar s$ quark flavor separation at large $x$ because of its sensitivity to different flavor combinations of valence-quark PDFs.
Experiments that will deliver new pion and kaon Drell-Yan data are also proposed for the CERN M2 beam line by the COMPASS++/AMBER collaboration \cite{Denisov:2018unj}, which would constrain the separated valence and sea quark pion PDFs above $x_\pi = 0.2$. The previously published HERA results on the pion Sullivan process would continue to be used to constrain the pion PDFs on $x_\pi < 10^{-3}$ at $Q^2 = 1\,$GeV$^2$ \cite{Barry:2018ort}.

It is anticipated that a combination of the global data set described above, used in conjunction with the anticipated EIC data, will form the basis for a rigorous meson parton distribution function analysis effort. Initial forays into extracting meson PDFs from data exist \cite{Wijesooriya:2005ir, Aicher:2010cb, Barry:2018ort} already. In all such efforts, parametric form, model, theory, and other uncertainties come into play, just as with the multi-decade, ongoing nucleon PDF extractions \cite{Kovarik:2019xvh}.  However, such considerations will not alter the salient features; and the single most important factor to reduce the meson PDF uncertainties is increased data, in particular, increased kinematic range of data.

Supposing, as accumulating evidence suggests, that a material nonzero domain in $-t$ exists whereupon one can extract physical $\pi$ ($K$) information using the Sullivan process, then within the projected EIC luminosity reach and detection capabilities, one could even envision measurements of the pion's GPD. Projected experimental results would be, at least, at the level achieved previously at HERA for the proton.  Moreover, should the data validate the assumption that reliable $\pi$ ($K$) structure data can be extracted for $-t \leq 0.6 (0.9)\,$GeV$^2$, then one would even gain an order-of-magnitude in statistics as compared to HERA proton data.

A strong motivation for such measurements is the fact that the leading $x$-moment of the pion's (kaon's) GPD provides access to the distributions of mass and momentum within the pion (kaon).  Since these distributions can be calculated \cite{Broniowski:2008hx, Frederico:2009fk, Mezrag:2014jka, Kumano:2017lhr, deTeramond:2018ecg, Shanahan:2018pib, Lan:2019vui}, such data can significantly influence future theoretical perspectives.  More speculatively, using elastic $J/\Psi$ and $\Upsilon$ production near threshold, one could attempt to obtain guidance on the trace anomaly contributions to the $\pi$ ($K$) mass, with uncertainties perhaps a factor of ten larger than those for the proton. At the same time, the data would provide the unique opportunity to gain valuable insights regarding the pion's transverse-momentum dependent structure at both leading and sub-leading twist \cite{Lorce:2016ugb}, and similarly for the kaon, from semi-inclusive deep inelastic scattering measurements.

\subsection{Gluon Energy in the Pion and Kaon}
Alternatively, one can directly access gluon PDFs in the pion and kaon through open charm production \cite{Shifman:1977yb} from the proton's meson cloud using the versatility of EIC, with its high luminosity and energy reach. In the chiral limit and at very high momentum transfer squared ($Q^2$) scales, the gluon distributions in the proton are expected to grow with $Q^2$ until a saturation scale is ultimately reached. Equally in the chiral limit, there remain puzzles for the pion, \emph{e.g}.\ do the gluons disappear in this limit in the pion, or do gluons persist and cancellations keep the pion near-massless?

Such open charm measurements would require large longitudinal momentum ($x_L$) scales; hence, excellent hadronic calorimetry energy resolution and sufficiently high luminosity to produce copious open charm events in the small-$t$ region. Ideally, EIC would provide a map of data in fine bins of $x_L$ and $-t$ and therewith convincing evidence that one can truly extract the gluon distributions of a physical pion and/or kaon at large $x$ $(\gtrsim 0.2)$, as a function of $Q^2$. On this domain, EIC could potentially provide measurements of gluon PDFs in the $\pi$ and $K$ over a range in $Q^2$ from a few GeV$^2$ to a few 100\,GeV$^2$, thus providing a good lever arm to map the $Q^2$-dependence and settle if gluons in the pion disappear or persist.

At large $x$, lQCD could also provide information on gluon distributions in pions and kaons as extracted from the lowest structure function moments. Such a lQCD calculation would have strong synergy with the EIC science program.  EIC could also provide the fractional gluon contributions to these moments at lower values of $x$, so that one could, in principle, use these as part of the calculated lattice moments to better constrain the extracted values at large-$x$ and compare with experimental EIC data.  lQCD could potentially also provide guidance on the gluon distributions in exotic/hybrid states and help develop physical intuition regarding their contribution to the masses of such states.  Steps in these directions are being taken \cite{Yang:2018bft}.

\subsection{Pion and Kaon Form Factors}
The elastic electromagnetic form factors of the charged pion and kaon, $F_\pi(Q^2)$ and $F_K(Q^2)$, are a rich source of insights into basic features of hadron structure, such as the roles played by confinement and DCSB in fixing the hadron's size, determining its mass, and defining the transition from the strong- to perturbative-QCD domains (see Fig.\,\ref{figradius} and the associated discussion). Studies during the last decade, based on JLab 6-GeV measurements, have generated confidence in the reliability of pion electroproduction as a tool for pion form factor extractions (see Sect.\,\ref{secPiKSullivan}). Forthcoming measurements at the 12-GeV JLab will deliver pion form factor data that are anticipated to bridge the transition region.  Starting in the long-distance (small-$Q^2$) domain, where $F_\pi(Q^2)$ is characterized by $r_\pi$, and moving to shorter distances (larger $Q^2$), where power-law scaling and logarithmic scaling violations, characteristic of hard gluon and quark degrees-of-freedom, become visible in the $Q^2$-dependence of $F_\pi(Q^2)$, but its magnitude is set by a light-front pion wave function which is dilated owing to DCSB \cite{Chang:2013pq}.  Both lQCD and continuum methods are being developed in order to provide robust predictions for the breadth and character of this transition domain \cite{Wang:2018pii, Chambers:2017tuf, Koponen:2017fvm, Chen:2018rwz}.

\begin{figure}[t]
\centerline{%
\includegraphics[clip, width=0.42\textwidth]{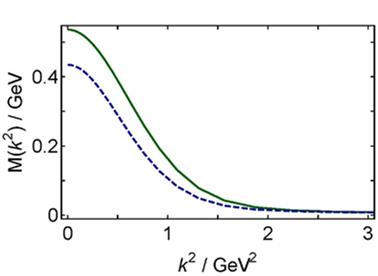}}
\centerline{%
\includegraphics[clip, width=0.42\textwidth]{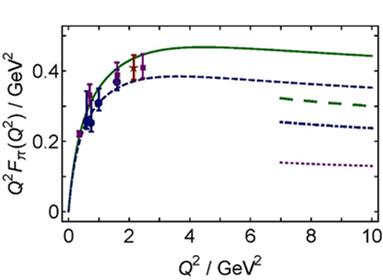}}
\caption{\label{figF9}
\emph{Upper panel}. Two dressed-quark mass functions distinguished by the amount of DCSB: emergent mass generation is 20\% stronger in the system characterized by the solid green curve, which describes the more realistic case. \emph{Lower panel}. $F_\pi(Q^2)$ obtained with the mass function in the upper panel: $r_\pi = 0.66\,$fm with the solid green curve and $r_\pi = 0.73\,$fm with the dashed blue curve. The long-dashed green and dot-dashed blue curves are predictions from the QCD hard-scattering formula, obtained with the related, computed pion PDAs. The dotted purple curve is the result obtained from that formula if the asymptotic profile is used for the PDA: $\varphi(x)=6x(1-x)$.
}
\end{figure}

The connection between such observables and mass generation in the Standard Model is illustrated in Fig.\,\ref{figF9} \cite{Chen:2018rwz}.
%\cite{Chen:2018rwz, MChenPrivate}.
The upper panel depicts two similar but distinct dressed light-quark mass-functions, characterized by a different DCSB strength, \emph{i.e}.\ the $k^2 = 0$ value.  The solid green curve was computed using a QCD effective charge whose infrared value is consistent with modern continuum and lattice analyses of QCD's gauge sector \cite{Binosi:2016nme, Rodriguez-Quintero:2018wma}, whereas the dashed blue curve was obtained after reducing this value by 10\%.  In a fully-consistent calculation, such a modification is transmitted to every element in the calculation, \emph{viz}.\ propagators, bound-state wave function, and photon-quark coupling; and subsequently to all observables. The resulting impact on $F_\pi(Q^2)$ is depicted in the lower panel: evidently, this experiment is a sensitive probe of the strength of emergent mass generation.  The lower panel of Fig.\,\ref{figF9} also depicts results obtained using the QCD hard scattering formula derived for pseudoscalar mesons \cite{Lepage:1979zb, Efremov:1979qk, Lepage:1980fj}.  As noted above and explained elsewhere \cite{Chang:2013nia, Gao:2017mmp}, at empirically accessible energy scales they, too, are sensitive to the emergent mass scale in QCD.

\begin{figure}[t]
\centerline{%
\includegraphics[clip, width=0.44\textwidth]{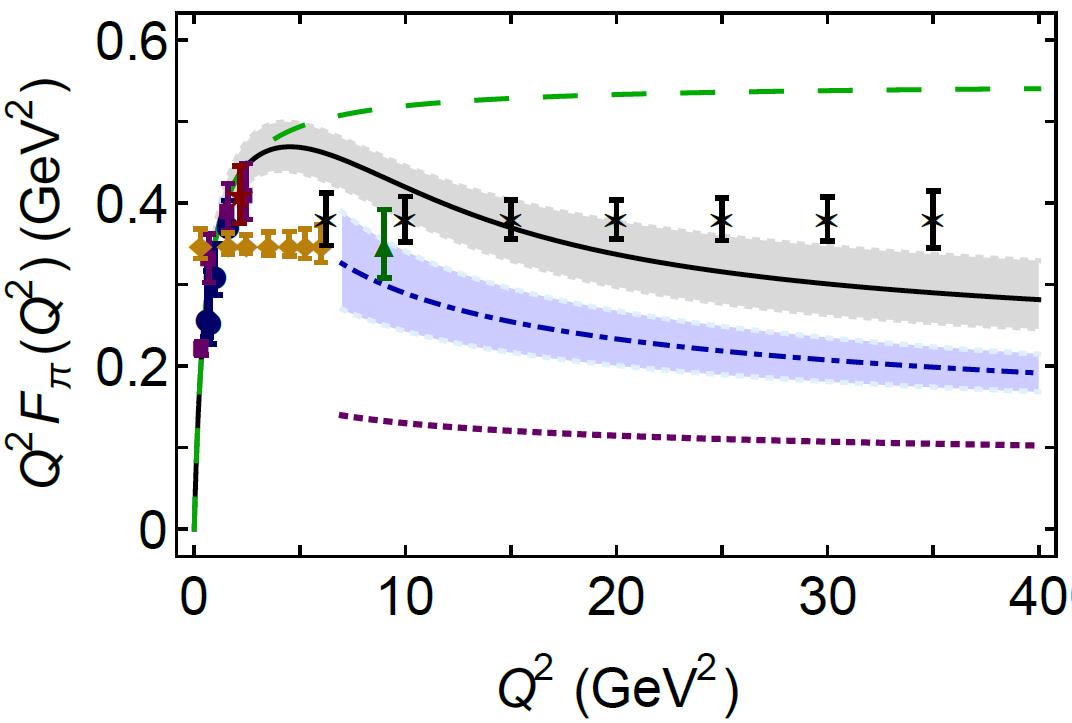}}
\caption{\label{figF10}
Projected EIC pion form factor data as extracted from a combination of electron-proton and electron-deuteron scattering, each with an integrated luminosity of $20\,{\rm fb}^{-1}$ -- black stars with error bars. Also shown are projected JLab 12-GeV data from a Rosenbluth-separation technique -- orange diamonds and green triangle. The long-dashed green curve is a monopole form factor whose scale is determined by the pion radius. The black solid curve is the QCD-theory prediction bridging large and short distance scales, with estimated uncertainty \cite{Chen:2018rwz}. The dot-dashed blue and dotted purple curves represent the short-distance views \cite{Lepage:1979zb, Efremov:1979qk, Lepage:1980fj}, comparing the result obtained using a modern DCSB-hardened PDA and the asymptotic profile, respectively.
}
\end{figure}

At EIC, pion form factor measurements can be extended to still larger $Q^2$, by measuring ratios of positively- and negatively-charged pions in quasi-elastic electron-pion (off-shell) scattering using the Sullivan process. The measurements would be over a range of small $-t$, constrained to sufficiently small virtuality and gauged with theoretical and phenomenological expectations, to again verify the reliability of the pion form factor extraction.

A consistent and robust EIC pion form factor data set will probe deep into the region where $F_\pi(Q^2)$ exhibits strong sensitivity to both emergent mass generation via DCSB and the evolution of this effect with scale. Figure~\ref{figF10} shows the EIC projections for possible pion form factor measurements. The pion form factor projections assume an integrated luminosity of $20\,{\rm fb}^{-1}$ with a $5\,$GeV electron beam colliding with a $100\,$GeV proton beam. The uncertainties include statistical and 2.5\% point-to-point and 12\% scale systematic uncertainties (similar to what was found for the HERA-H1 pion structure function data), and are propagated from the cross-section from which the pion form factor is extracted.  Moreover, since a longitudinal/transverse separation of the cross-section is not possible at EIC, owing to an inability to access small photon polarization; the uncertainty owing to the model used to estimate the longitudinal and transverse contributions to the cross-section is also considered. The model is assumed to have been validated by the ratio of charged-pion $\pi^+/\pi^-$ data extracted separately from electron-proton and electron-deuteron beam collisions with equivalent center-of-mass energy and integrated luminosity.

Measurements at the 12-GeV JLab on exclusive kaon electroproduction beyond the resonance region at $-t \leq 0.9\,$GeV$^2$ and $Q^2$ up to $\approx 5\,$GeV$^2$ will need to be completed to provide similar guidance on whether high-$Q^2$ kaon form factor measurements can be feasible at EIC.

\subsection{Valence-Quark Distributions in the Pion and Kaon at Large Momentum Fraction $\mathbf x$}
\label{SecKaonStructure}
As noted above (see Fig.\,\ref{figPDAs}), comparisons between distributions of truly light quarks and those describing $s$-quarks may be ideally suited to exposing measurable signals of dynamical mass generation. A striking example can be found in a comparison between the valence-quark PDFs of the pion and kaon at large $x$. A significant disparity between these distributions would point to a marked difference between the fractions of pion and kaon momentum carried by the other bound state participants, particularly gluons. Both phenomenological observations (\emph{e.g}.\ a heavier quark should radiate less gluons than a light quark) and continuum QCD theory \cite{Chen:2016sno} predict that the gluon content of the pion is vastly greater than that of the kaon.\footnote{Developments in lQCD theory and practice mean that contemporary simulations can potentially validate this projection \cite{Xu:2018eii}.}

This may be viewed as an expression of the near-complete cancellation in the almost massless pion between strong-mass-generating dressing of the valence-quark and -antiquark on one hand and binding attraction on the other.  Such effects set this system apart from the more massive kaon, wherein the cancellation is much less effective owing to the larger value of the Higgs-generated $s$-quark current-mass. Consequently, high-precision measurements of the valence-quark distributions in the pion and kaon at large $x$, over a range of momentum transfer scales that will enable guidance to be drawn for a rigorous QCD interpretation, are a high-priority; and they are achievable with the EIC.

\begin{figure}[t]
\centerline{%
\includegraphics[clip, width=0.42\textwidth]{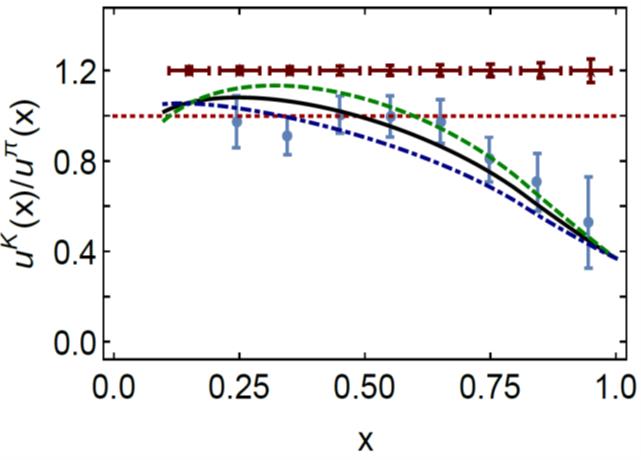}}
\vspace*{-2ex}

\caption{\label{figF11}
Ratio of valence $u$-quark PDFs in the pion and the kaon at $\zeta = 5.2\,$GeV$=:\zeta_5$.
Data are from Drell-Yan measurements \cite{Badier:1980jq}.
The computed results are taken from Ref.\,\cite{Chen:2016sno}, with the dashed, solid, and dot-dashed curves representing, respectively, $0$, $5$\%, $10$\% of the kaon's light-front momentum carried by glue at the scale, $\zeta_K = 0.51\,$GeV.  For the projected EIC data (brown points drawn at $u_K(x)/u_\pi(x)=1.2$) we assumed $u$-quark dominance.
(For reference, the horizontal dotted line marks $u_K(x)/u_\pi(x)=1$.)
}
\end{figure}

The quality of possible EIC measurements is illustrated in Fig.\,\ref{figF11}; evidently, they can effectively determine the anticipated differences between the pion and kaon. These measurements at large $x$ will drive the hadronic calorimetry requirements to pinpoint the scattering kinematics, whereas all described measurements require good angular or transverse momentum resolution and complete kinematic coverage.

Such planning is complemented by ongoing progress in theory.
Marking one significant class of advances, novel lQCD algorithms \cite{Liu:1993cv, Ji:2013dva, Radyushkin:2016hsy, Radyushkin:2017cyf, Chambers:2017dov} are beginning to yield results for the pointwise behavior of the pion's valence-quark distribution \cite{Chen:2018fwa, Karthik:2018wmj, Karpie:2019eiq, Sufian:2019bol,  Izubuchi:2019lyk}, offering promise for information beyond the lowest few moments \cite{Best:1997qp, Detmold:2003tm, Brommel:2006zz, Oehm:2018jvm}.
Extensions of the continuum analysis in Ref.\,\cite{Hecht:2000xa} are also yielding new insights.  For example,
a class of corrections to the handbag-diagram representation of the virtual-photon--pion forward Compton scattering amplitude has been identified and shown to restore basic symmetries in calculations of valence-quark distribution functions \cite{Chang:2014lva}.  (Such corrected expressions were used in Ref.\,\cite{Chen:2016sno}.)

Capitalising on these new developments, a recent, parameter-free continuum analysis delivered predictions for the valence, glue and sea distributions within the pion \cite{Ding:2019lwe}; unifying them with, \emph{inter alia}, electromagnetic pion elastic and transition form factors.  The analysis reveals that the valence-quark distribution function is hardened by DCSB and produces the following apportioning of momentum at the scale $\zeta=\zeta_2$:
\begin{subequations}
\label{momfractions}
\begin{align}
\langle x_{\rm valence} \rangle &= 0.48(3)\,,\\
\langle x_{\rm glue} \rangle  &= 0.41(2)\,,\\
\langle x_{\rm sea} \rangle  &= 0.11(2)\,.
\end{align}
\end{subequations}
These results are consistent with a phenomenological analysis of data on $\pi$-nucleus Drell-Yan and leading neutron electroproduction \cite{Barry:2018ort}: $\langle 2 x \rangle_{u}^\pi  = 0.48(1)$ at $\zeta=2.24\,$GeV.  Moreover, the glue and sea ordering agrees with that in \cite{Barry:2018ort}: in detail, the gluon momentum-fraction in Eq.\,\eqref{momfractions} is $\sim 20$\% larger and that of the sea is commensurately smaller.

\begin{figure}[t]
\centerline{%
\includegraphics[clip, width=0.44\textwidth]{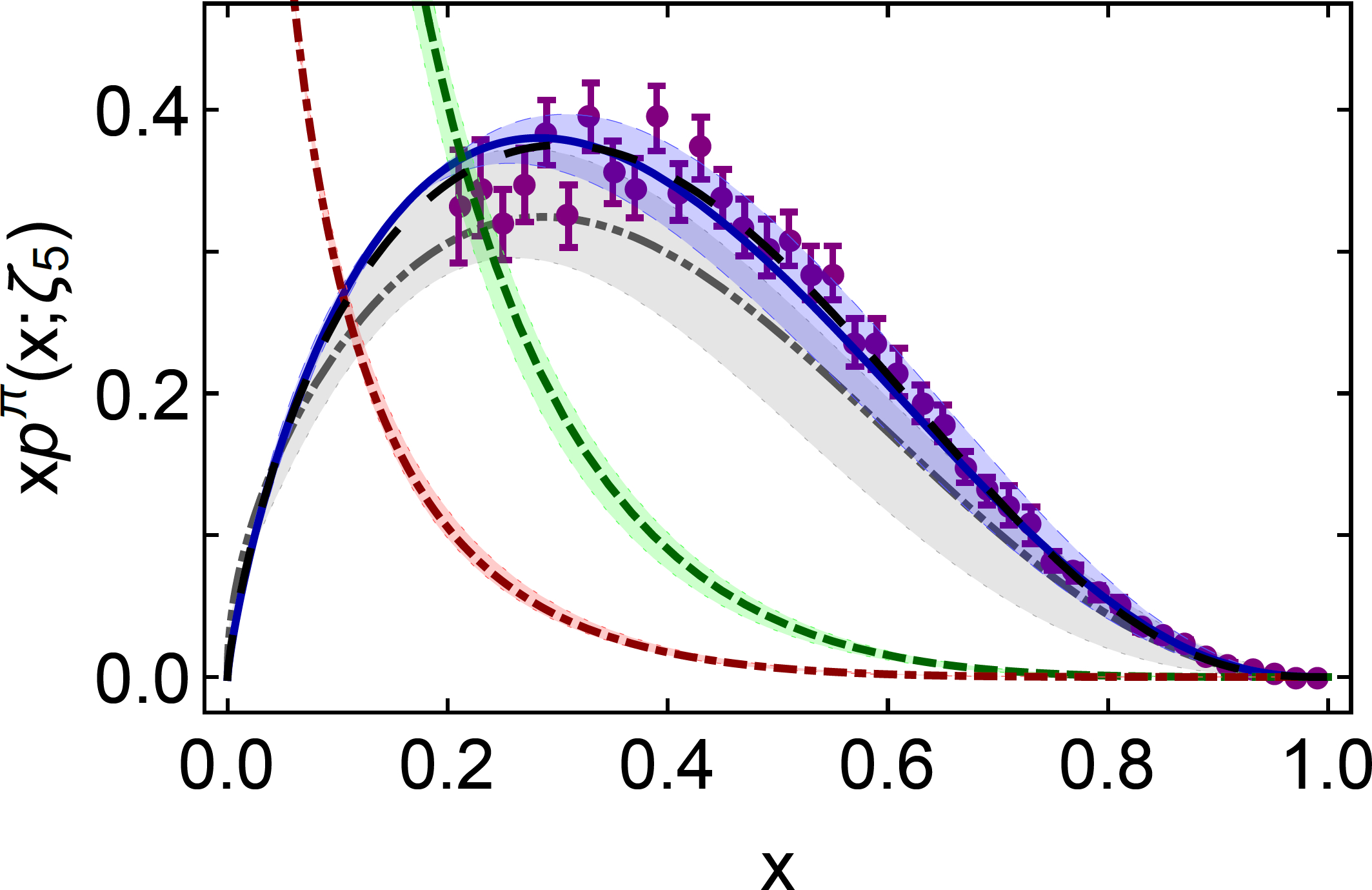}}
\caption{\label{figF12}
Pion valence-quark momentum distribution function, $x {\mathpzc u}^\pi(x;\zeta_5)$:
dot-dot-dashed (grey) curve within shaded band -- lQCD result \cite{Sufian:2019bol};
long-dashed (black) curve -- early continuum analysis \cite{Hecht:2000xa};
and solid (blue) curve embedded in shaded band -- modern, continuum calculation \cite{Ding:2019lwe}.
Gluon momentum distribution in pion, $x g^\pi(x;\zeta_5)$ -- dashed (green) curve within shaded band;
and sea-quark momentum distribution, $x S^\pi(x;\zeta_5)$ -- dot-dashed (red) curve within shaded band.
(In all cases, the shaded bands indicate the size of calculation-specific uncertainties, as described elsewhere \cite{Ding:2019lwe}.)
Data (purple) from Ref.\,\cite{Conway:1989fs}, rescaled according to the analysis in Ref.\,\cite{Aicher:2010cb}.}
\end{figure}

Importantly, as illustrated in Fig.\,\ref{figF12}, after evolution to $\zeta=\zeta_5$, the continuum prediction for $u^\pi(x)$ from Ref.\,\cite{Ding:2019lwe} matches that obtained using lQCD \cite{Sufian:2019bol}.  Given that no parameters were varied in order to achieve this or any other outcome in Ref.\,\cite{Ding:2019lwe}, one has arrived at a remarkable, modern confluence, which suggests that real strides are being made toward understanding pion structure.   It is essential to extend these continuum- and lattice-QCD studies to the analogous problem of predicting kaon distribution functions.

\subsection{Quark Fragmentation into Pions or Kaons}
Central to solving QCD is an elucidation of the nature of confinement and its connection with DCSB. In the presence of light quarks, confinement is a dynamical process: a gluon or quark is produced and begins to propagate in spacetime; but after a short interval, an interaction occurs so that the parton loses its identity, sharing it with others. Finally, a cloud of partons is produced, which coalesces into color-singlet final states. This is the physics of parton fragmentation functions (PFFs), which describe how QCD partons, generated in a high-energy event and (nearly) massless in perturbation theory, convert into a shower of massive hadrons, \emph{i.e}. they describe how hadrons with mass emerge from massless partons.

\begin{figure}[t]
\centerline{%
\includegraphics[clip, width=0.42\textwidth]{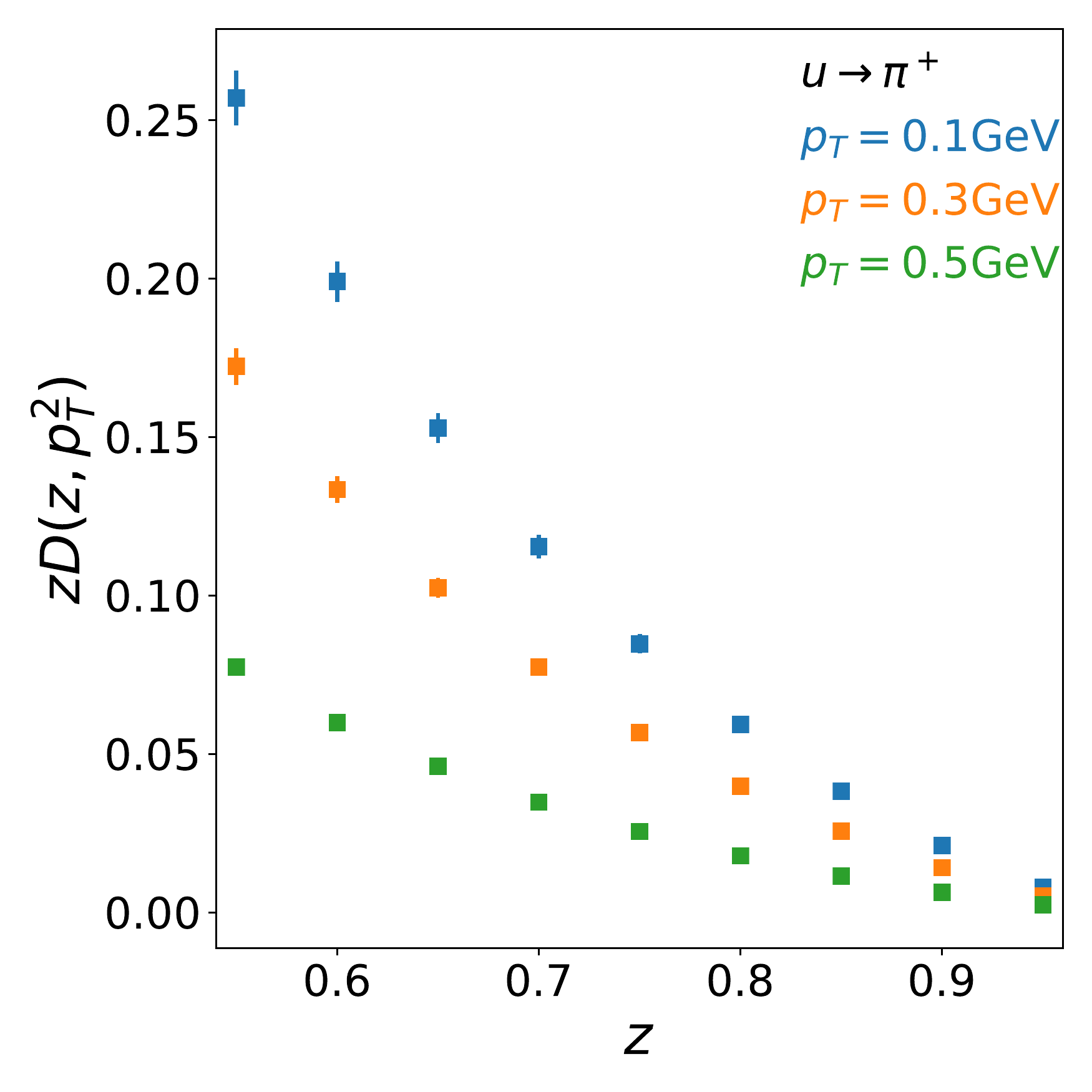}}
\centerline{%
\includegraphics[clip, width=0.42\textwidth]{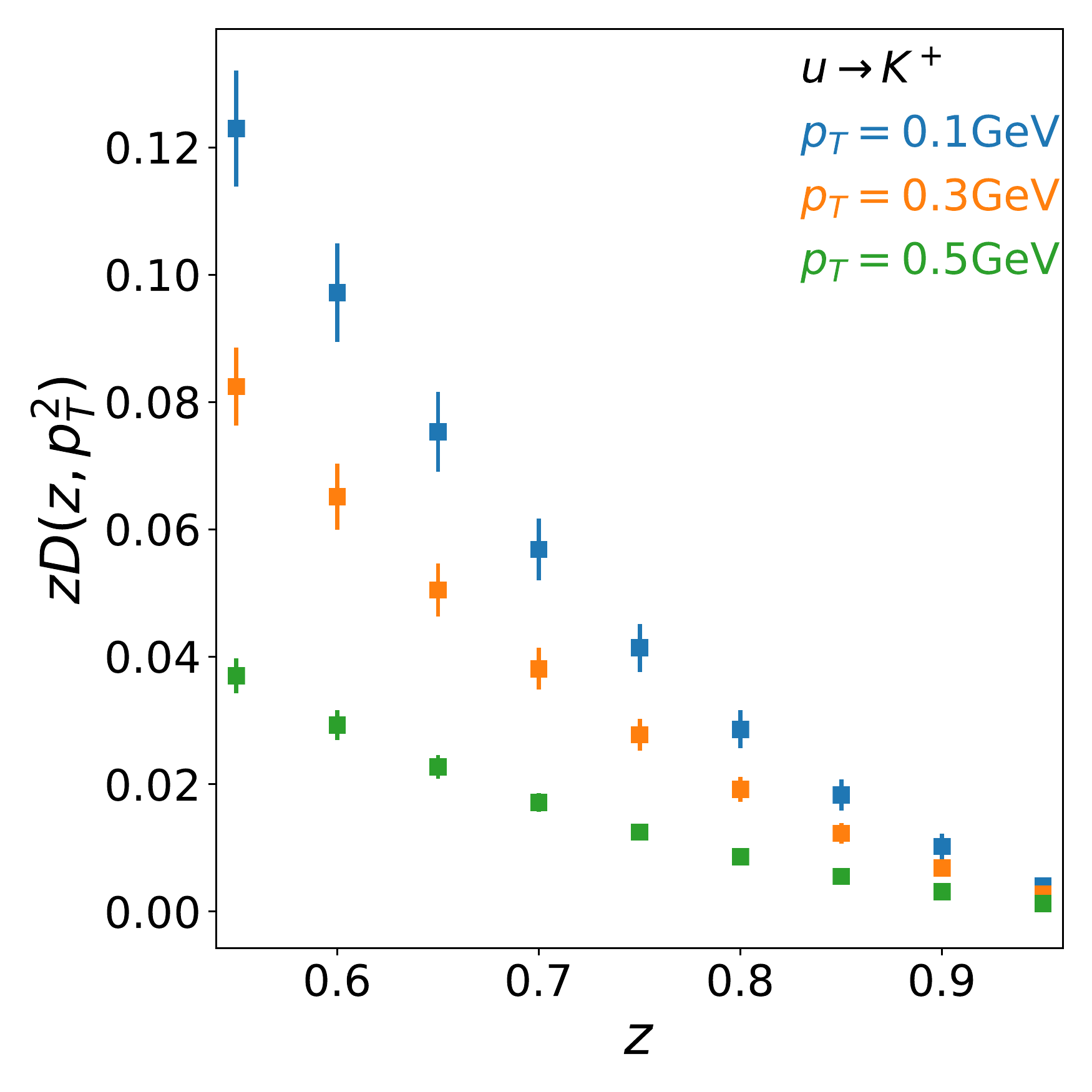}}
\caption{\label{figF13}
Projected uncertainties for measurements of the $u$-quark to pion (\emph{upper panel}) and kaon (\emph{lower panel}) fragmentation function at EIC for an integrated luminosity of $10\,{\rm fb}^{-1}$, for the large $z$ region, $z > 0.5$, and transverse momentum $k_\perp$ (as picked up in the fragmentation process) of $k_\perp = 0.1, 0.3, 0.5\,$GeV, respectively.
}
\end{figure}

Such observations support a view that PFFs are the cleanest expression of dynamical confinement in QCD. Moreover, owing to Gribov-Lipatov reciprocity \cite{Gribov:1971zn},  PDFs and PFFs are related by crossing symmetry in the neighborhood of their common boundary of support. Hence, like PDFs, PFFs provide basic insights into the origin of mass, serving as timelike analogs and providing a basic counterpoint to the PDFs.

In addition, every cross-section that can yield a given hadronic transverse-momentum-dependent parton distribution (TMD = 3D momentum image) involves a related PFF and requires knowledge of their dependence on both $z$ and $k_\perp$, where $z$ is the fractional energy carried away by the hadron and $k_\perp$ is the transverse momentum acquired. Evidently, the future of momentum imaging depends critically on making significant progress with the measurement, computation and understanding of PFFs. Notwithstanding these demands, there are currently no realistic computations of PFFs. Indeed, even a formulation of the problem remains uncertain.

EIC will provide precision data on quark fragmentation into a pion or kaon as a function of $(z, k_\perp)$, at large $z$ ($> 0.5$) and small $k_\perp$ ($< 1\,$GeV), as illustrated in Fig.\,\ref{figF13}. This is a crucial domain, testing most directly those aspects of QCD calculations that incorporate and express emergent phenomena, such as confinement, DCSB, and bound-state formation. Indeed, the fact that only bound-states emerge from such collisions is one of the cleanest available manifestations of confinement. EIC will be unique in these measurements as compared to fixed-target experiments, since one needs the energy range, versatility, and excellent detection capabilities in the collider environment to first cleanly single out the pion or kaon target and subsequently the fragmentation process tag. The empirical effort will deliver multi-dimensional bins in $x$, $Q^2$, $z$, and transverse hadron momentum, $P_T$, as desired by both phenomenology, for fitting and use in the analysis of a diverse array of processes, and theory aimed at the calculation and interpretation of PFFs.

\section{Conclusion}
A striking feature of the strong interaction is its emergent $1$-GeV mass-scale, as exhibited in the masses of protons, neutrons and numerous other everyday hadronic bound states. In sharp contrast, the energy associated with the gluons and quarks confined inside the strong interaction's Nambu-Goldstone bosons, such as the pion and kaon, is not so readily apparent. Even if both quarks and gluons acquire mass dynamically, in all hadrons, the pion ends up near-massless, and the kaon, where Higgs-driven and emergent mass generating mechanisms compete, ends up acquiring just half the $1$-GeV mass scale.

At EIC, pion and kaon structure can be measured through the Sullivan process, which necessarily means mesons are accessed off-shell. Nevertheless, recent experimental and phenomenological work strongly indicates that, under certain achievable kinematic conditions, the Sullivan process provides reliable access to a true meson target. Moreover, the off-shell dependence is measurable. At EIC, measurements could be made as a function of $-t$, and the kinematic conditions required to obtain physical pion and kaon structure information can be charted and exploited further.

Herein, we have identified and described five key EIC measurements that can be expected to deliver far-reaching insights into the dynamical generation of mass, the crucial feature of the Standard Model which leads to apparently mysterious differences between pion, kaon and proton structure.
\begin{enumerate}[label=(\roman*)]
\item Measurement of pion and kaon structure functions and their generalized parton distributions will render insights into quark and gluon energy contributions to hadron masses.
\item Measurement of open charm production will settle the question of whether gluons persist or disappear within pions in the chiral limit -- if they persist it proves the cancellation of terms that must occur such that the pion mass is driven by Higgs-generated current quark masses, albeit with a huge emergent magnification factor.
\item Measurement of the charged-pion form factor up to $Q^2 \approx 35\,{\rm GeV}^2$, which can be quantitatively related to emergent-mass acquisition from dynamical chiral symmetry breaking.
\item Measurement of the behavior of (valence) $u$-quarks in the pion and kaon, which gives a quantitative measure of the contributions of gluons to NG boson masses and differences between the impacts of emergent and Higgs-driven mass generating mechanisms.
\item Measurement of the fragmentation of quarks into pions and kaons, a timelike analog of mass acquisition, which can potentially reveal relationships between dynamical chiral symmetry breaking and the confinement mechanism.
\end{enumerate}

\begin{acknowledgements}
The structure of the science program outlined herein emerged from discussions during two dedicated ``Workshops on Pion and Kaon Structure at an Electron-Ion Collider''.  The first, PIEIC2017 -- \href{http://www.phy.anl.gov/theory/pieic2017/}{http://www.phy.anl.gov/theory/pieic2017/}
took place at Argonne National Laboratory, 1-2 June 2017; and the second, PIEIC2018 -- \href{https://www.jlab.org/conferences/pieic18/}{https://www.jlab.org/conferences/pieic18/},
was held at the Catholic University of America, 24-25 May 2018.
The development of this document was also informed by discussions at later meetings, such as the ``Workshop on Emergent Mass and Its Consequences in the Standard Model'', at the European Centre for Theoretical Studies in Nuclear Physics and Related Areas, \href{https://indico.ectstar.eu/event/23/}{https://indico.ectstar.eu/event/23/}, Trento, Italy, 17-21 September 2018.
We acknowledge constructive input from all participants; and are grateful to the following people for valuable comments during the preparation of this manuscript:
Z.-F.~Cui,
G. Eichmann,
C.\,S.~Fischer,
A.~Freese,
P.~Hoyer,
B.-L.~Li,
Y.-X.~Liu,
Y.~Lu,
W.~Melnitchouk,
J.~Pawlowski,
O.~Philipsen,
G.~Salme,
C.~Shi,
B.~Wojtsekhowski,
P.-L.~Yin,
and
H.-S.~Zong.
%% Add all workshop participants.
%
Work supported by:
Alexander von Humboldt Foundation through a Postdoctoral Research Fellowship;
China's \emph{Thousand Talents Plan for Young Professionals};
Consejo Nacional de Ciencia y Tecnolog\'ia (CONACyT), Mexico, Grant Nos.\ 4.10 and CB2014-22117;
Conselho Nacional de Desenvolvimento Cient\'{\i}fico e Tecnol\'ogico - CNPq, Grants No.\
305894/2009-9,
464898/2014-5,
305815/2015 and
308486/2015-3
(INCT F\'{\i}sica Nuclear e Apli\-ca\-\c{c}\~oes);
Coordinaci\'on de la Investigaci\'on Cient\'ifica (CIC) of the University of Michoac\'an;
Deutsche Forschungsgemeinschaft, Grant No.\ FI 970/11-1;
Forschungszentrum J\"ulich GmbH;
Funda\c{c}\~ao de Amparo \`a Pesquisa do Estado de S\~ao Paulo (FAPESP), Grant Nos.\
2013/01907-0,
2017/05660-0,
2017/05685-2 and
2017/07595-0;
Fundamental Research Funds for the Central Universities (China), under Grant No.\ 2019CDJDWL0005; % Sixue
Helmholtz International Centre for FAIR within the LOEWE program of the State of Hesse;
Jiangsu Province \emph{Hundred Talents Plan for Professionals};
Junta de Andaluc\'ia Grant No.\ FQM-370; % Jorge
Nanjing University of Posts and Telecommunications Science Foundation, under Grant No.\ NY129032; %Shu-Sheng
Natural Science Foundation of China, under Grant Nos.\ 11805024, 11847024 and 11847301;  % Sixue & Shu-Sheng
Natural Sciences and Engineering Research Council of Canada (NSERC), under Grant No.\ SAPIN-2016-00031;
Spanish MEyC under Grant Nos.\ FPA2017-84543-P and SEV-2014-0398; % Joannis
Spanish MINECO Grant No.\ FPA2017-86380-P; % Jorge
Spanish National Project FPA2017-86380-P; % Pepe
U.S.\ Department of Energy, Office of Science, Office of Nuclear Physics, under contract nos.\,DE-AC02-06CH11357, DE-AC05-06OR23177 and DE-FG02-87ER40371;
% Roberts ... Mokeev
%
and
U.S.\ National Science Foundation, Grant Nos.\ PHY-1506416, PHY-1714133 and PHY-1812382. % V. Andrieux, Ralf Gothe
\end{acknowledgements}

%%\bibliographystyle{../../../zProc/z10/z10KITPC/h-physrev4}
%%\bibliography{../../../CollectedBiB}

\end{document}